\documentclass[journal]{IEEEtran}
\usepackage{multirow}
\usepackage{cite}
\usepackage{graphicx}
\usepackage{stfloats}
\usepackage{subfigure}
\usepackage[utf8]{inputenc}
\usepackage[T1]{fontenc}
\usepackage{color}
\usepackage{soul}
\usepackage{alltt}
\usepackage{float}
\usepackage{epstopdf}
\usepackage{enumerate}
\usepackage{bm}
\usepackage{amsmath}
\usepackage{makecell}
\usepackage{newunicodechar}
\usepackage{amsfonts,amssymb}
\usepackage{tabularx}
\usepackage{threeparttable}
\usepackage{caption}
\usepackage{authblk}
\usepackage[justification=centering]{caption}
\newunicodechar{ﬁ}{fi}
\newunicodechar{ﬀ}{ff}

\DeclareUnicodeCharacter{FF0C}{-}
\DeclareUnicodeCharacter{3001}{-}
\DeclareUnicodeCharacter{04BB}{-}
\DeclareUnicodeCharacter{A3}{-}
\DeclareUnicodeCharacter{00A0}{~}
\begin{document}


\title{Graph neural network-based fault diagnosis: a review}

\author{
	\vskip 1em
	{\color{black}
	Zhiwen Chen, Jiamin Xu, Cesare Alippi, Steven X. Ding, Yuri Shardt, Tao Peng,  Chunhua Yang
	}

	\thanks{
		
		{\color{black}
		
		Zhiwen Chen, Jiamin Xu, Tao Peng and Chunhua Yang are with the School of Automation, Central South University, Changsha, 410083, China. Zhiwen Chen and Tao Peng are also with the Peng Cheng Laboratory, Shenzhen 518066, China.

        Cesare Alippi is with Universita della Svizzera italiana, Lugano, Switzerland, and Politecnico di Milano, Milano, Italy.	

		Steven X. Ding is with the Institute for Automatic Control and Complex Systems (AKS), University of Duisburg-Essen, Duisburg, 47057, Germany. 

        Yuri Shardt is with Institute of Automation and Systems Engineering, Technical University of Ilmenau, Ilmenau, Germany.

Email addresses of corresponding authors: zhiwen.chen@csu.edu.cn; cesare.alippi@polimi.it).
		}
	}
}


\maketitle

\begin{abstract}
Graph neural network (GNN)-based fault diagnosis (FD) has received increasing attention in recent years, due to the fact that data coming from several application domains can be advantageously represented as graphs. Indeed, this particular representation form has led to superior performance compared to traditional FD approaches. In this review, an easy introduction to GNN, potential applications to the field of fault diagnosis, and future perspectives are given. First, the paper reviews neural network-based FD methods by focusing on their data representations, namely, time-series, images, and graphs. Second, basic principles and principal architectures of GNN are introduced, with attention to graph convolutional networks, graph attention networks, graph sample and aggregate, graph auto-encoder, and spatial-temporal graph convolutional networks. Third, the most relevant fault diagnosis methods based on GNN are validated through the detailed experiments, and conclusions are made that the GNN-based methods can achieve good fault diagnosis performance. Finally, discussions and future challenges are provided.
\end{abstract}


\definecolor{limegreen}{rgb}{0.2, 0.8, 0.2}
\definecolor{forestgreen}{rgb}{0.13, 0.55, 0.13}
\definecolor{greenhtml}{rgb}{0.0, 0.5, 0.0}

\tableofcontents
\addcontentsline{file}{section_unit}{entry}

\section{Introduction}

In industry, fault diagnosis (FD) is required to guarantee safe and specification compliant operation of plants. This allows the process to attain industrial intelligence. Research on FD has a long history, and is also an important part of some related techniques, like condition monitoring \cite{ConMo, Lei_14}, process monitoring \cite{Yin_14,GE201716,ZHAO_Zhou_2021}, prognostic and health management (PHM) \cite{LEE2014314}, and abnormality management \cite{Wang_chenTW16,AbnMan}. Fault diagnosis methods can be divided into two categories, namely, model-based and data-based methods \cite{review,3}. Early fault diagnosis methods mainly belong to the model-based class, where the physical model or a state observer for the system of interest is constructed, and fault diagnosis is achieved by inspecting the changes in the residuals \cite{1,2,Alippi,Luo,ACTA1}.

In recent years, with the progress of sensor technology and the improvement of data storage capacity, data-driven fault diagnosis methods have becoming a research focus, with neural network (NN)-based method an important part due to powerful data processing capabilities. Researchers have developed a variety of NN architectures, such as convolutional neural network (CNN) architectures \cite{4}, residual network (ResNet) architectures \cite{5}, and Bayesian neural network (BNN) architectures \cite{6}. Inspired by these, researchers began to apply NN and deep learning models to the fault diagnosis field \cite{IFD,KPI,ACTA2,ACTA3,Yuri,ChenHT_20_review}. For example, H. Chen et al. \cite{ChenHT_20_review} reviewed the-state-of-the-art data driven methods with applications to high-speed trains. Besides, in \cite{10-}, a comprehensive review of the general architecture and principles of one-dimensional convolutional neural networks (1dCNNs) along with their main engineering applications was presented. M. Kuppusamy et al. \cite{11-} made a review on the application of deep learning techniques in five critical electrical applications. In \cite{FD_Survey}, three popular deep learning algorithms were briefly introduced, and their applications were reviewed through publications and research works on the area of bearing fault diagnosis.


The research and practice of NNs including CNN have proved outstanding results in numerous applications. However, in some research fields involving non-Euclidean structured data, the widely-used CNN models are not able to achieve optimal performance limited by its structural characteristics. For example, in \cite{43}, it is pointed out that papers refer to each other and construct a non-Euclidean graph structure citation network. In bio-engineering, predicting the functional types of protein based on its structure is a fascinating research field, while the internal structure of protein data can be abstracted as a complex graph structure \cite{14}. Due to the limitation of their network architecture and computation standard, CNNs cannot exploit existence of non-Euclidean graph structure, thus limits the application of CNNs.

\begin{figure}[tb]
\begin{center}
\includegraphics[scale=0.662]{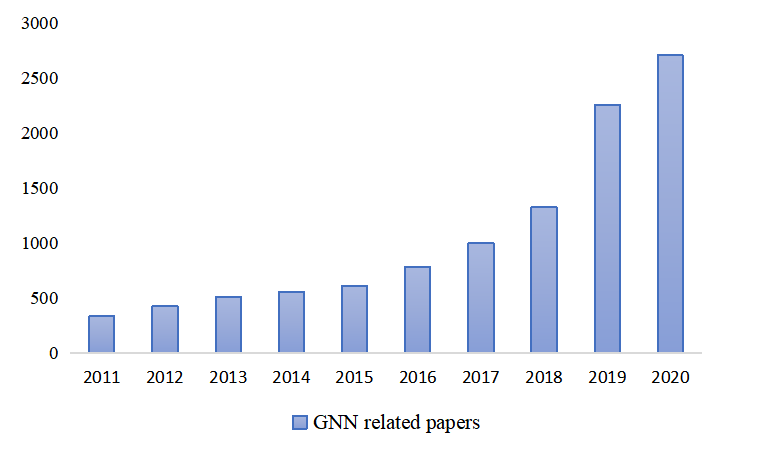}
\caption{Publication trend of GNN}\label{fig_1}
\end{center}
\end{figure}

\begin{figure}[tb]
\centering
\includegraphics[scale=0.268]{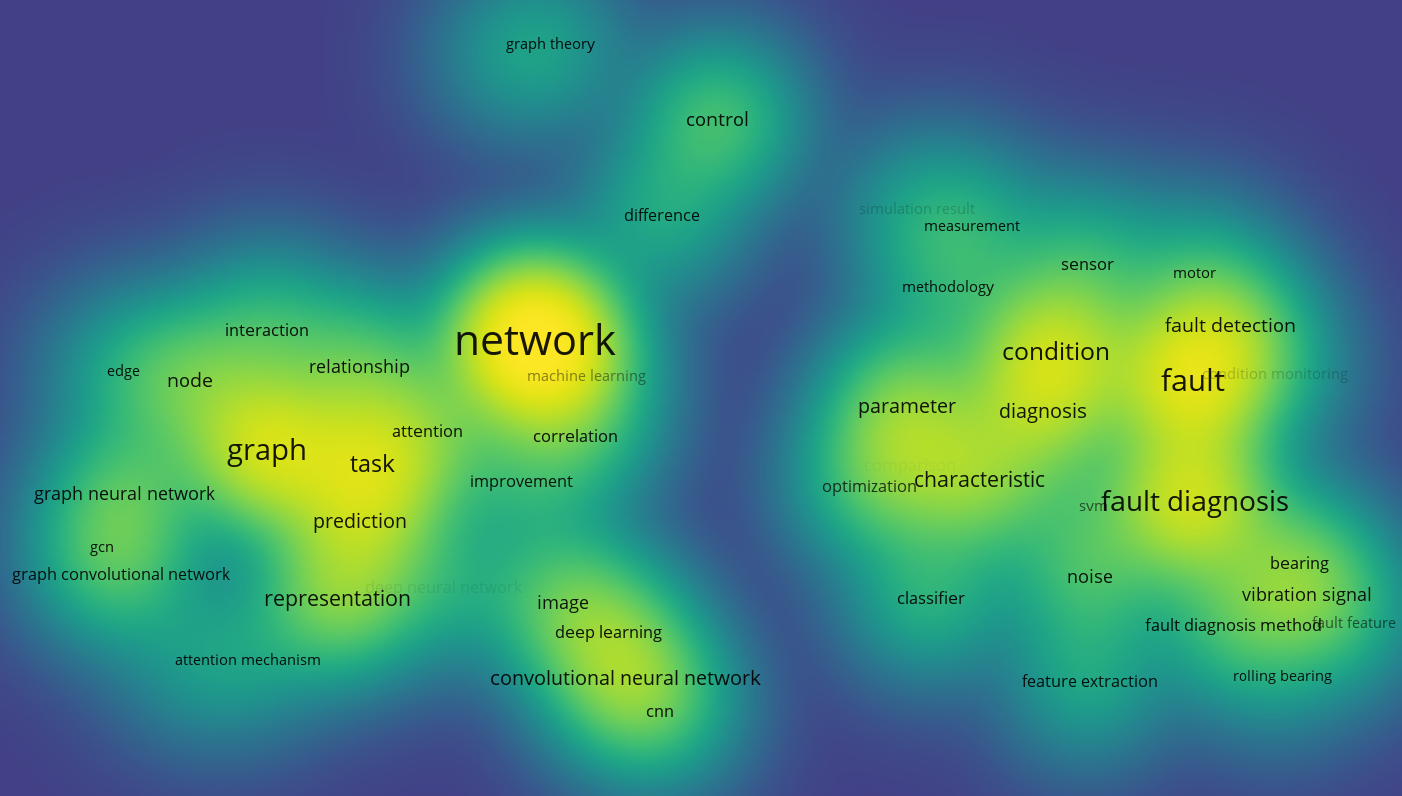}
\caption{Heat map of GNN related keywords}\label{fig_2}
\end{figure}

Graph neural networks (GNNs) were proposed to take advance of inductive bias associated with functional dependencies and, as such, exploit non-Euclidean representations \cite{15,16}. GNNs provide architectures inspired by the deep ones and offer suitable operators to process such information-rich structures \cite{R-CNN}. In fact, compared with deep neural network, GNN can process data characterized by complex spatiotemporal relationships. As a result, GNNs are widely used in computer vision \cite{17}, text processing \cite{18}, knowledge mapping \cite{19}, and recommendation systems. Fig. \ref{fig_1} records the number of papers related to GNN obtained from the Web of Knowledge using keyword ``Graph neural network''. It shows how the number of publications related to GNN is growing exponentially. Furthermore, Fig. \ref{fig_2} lists some hot words related to GNN and fault diagnosis, which are obtained by searching Web of Knowledge with keywords ``Graph neural network'' and ``Fault diagnosis''. The figure shows a large intersection between both research fields.

At present, there have been numerous review papers related to GNN \cite{GNNSur1,GNNSur2,GNNSur3}. In particular, existing GNN models were reviewed in \cite{20}, and several issues for future work raised. Later, \cite{22} investigated GNNs and the attention mechanism. In \cite{21}, a new taxonomy of GNN architectures was proposed, and the application of GNN to various fields was discussed.

In a real world, faults may change the running state of the process, and lead to a change of the dependency between measurements. Based on this understanding, researchers began to explore GNN for fault diagnosis tasks. For instance, in \cite{23}, a new graph convolutional network (GCN) framework for distribution network fault location was proposed, which not only obtains high model performance, but also shows good robustness properties. To deal with the problem of limited labeled data collected by an electromechanical system, \cite{24} constructed a semisupervised graph convolutional depth confidence network and an intelligent fault diagnosis method. In the field of fault diagnosis of wind turbine gearbox, based on the fact that the deep learning method cannot make full use of the relevant information of data, \cite{63} proposed a new fast deep GCN, and achieved improved performance. In \cite{25}, authors discussed some structural defects of GCN models, and proposed a multireceiving field graph convolutional network to realize effective intelligent fault diagnosis. \cite{26} proposed a new bearing fault diagnosis model based on horizontal visibility graph and GNN. However, most of these publications only focus on a small scope of research and seek to leverage GNN to a certain level only.

With that in mind, this review is prepared in response to the urgent need of a seamless and rigorous transition from classical neural network-based fault diagnosis to the alternative fault diagnosis methods which operate directly on graph-structured data. This paper focuses on investigating and reviewing GNN for fault diagnosis purpose. GNN-based fault diagnosis methods are studied and analyzed from multiple perspectives. The contributions of this paper are:


1. \textbf{New Taxonomy}. Neural network-based fault diagnosis methods are divided into three categories according to the representations of input data.

2. \textbf{Review-Oriented}. Several architectures of GNNs are reviewed, and the feasible applications of these architectures on fault diagnosis are explained. It is pointed out that one of difficulties of GNN-based fault diagnosis method lies in the construction of the association graph, so several feasible solutions are provided.

3. \textbf{Benchmark Study}. The diagnosis performance of several GNN-based methods and baseline methods are compared through three benchmark data sets. Based on the results obtained, a discussion follows.

4. \textbf{Future Research}. With the expectation of identifying possible research directions, several challenges related with GNN-based fault diagnosis are presented and discussed.

5. \textbf{Open Source}. The source code for this review will be made available after the peer-review stage. The code provides the implementation details of the GNN-based fault diagnosis methods discussed in this paper.

\section{Data-driven fault diagnosis}

\begin{figure*}
\centering
\includegraphics[scale=0.458]{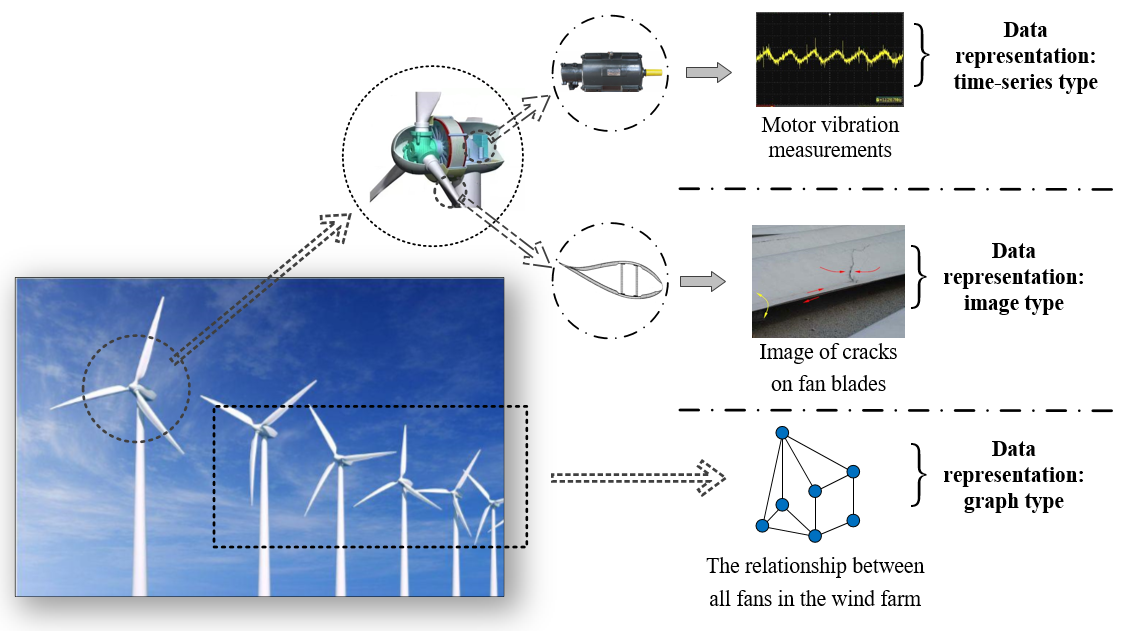}
\caption{Schematic diagram of data representation (sample example)}\label{fig_3}
\end{figure*}

With the rapid development of advanced sensing technologies and computational power, neural-network-based fault diagnosis (NNFD) has received considerable attention in industry. According to the various types of data representations, conventional NNFD methods can be divided into two categories, namely, image-based NNFD methods and time-series-based NNFD methods.

\begin{figure*}
\centering
\includegraphics[scale=0.458]{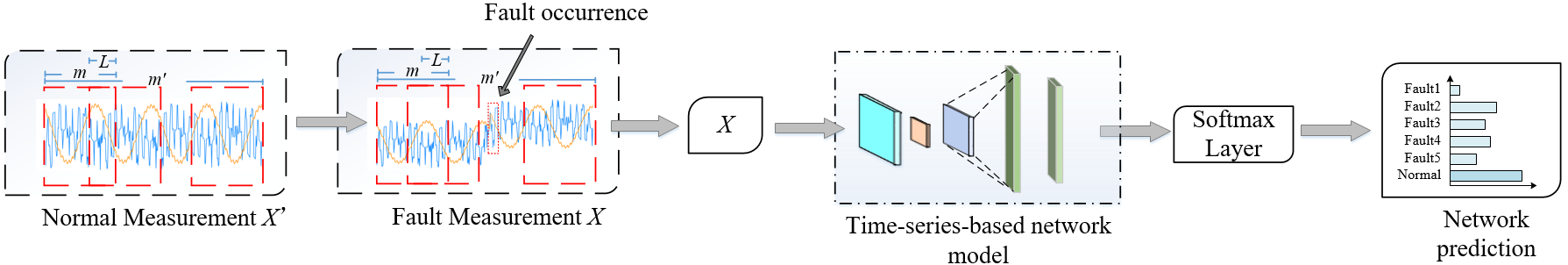}
\caption{Fault diagnosis framework based on time-series processing. A time-series-based network model is built to compute the latent representations of data. To compute the probability for fault categories, linear and a softmax layers are used.}\label{fig_4}
\end{figure*}

\begin{figure*}
\centering
\includegraphics[scale=0.501]{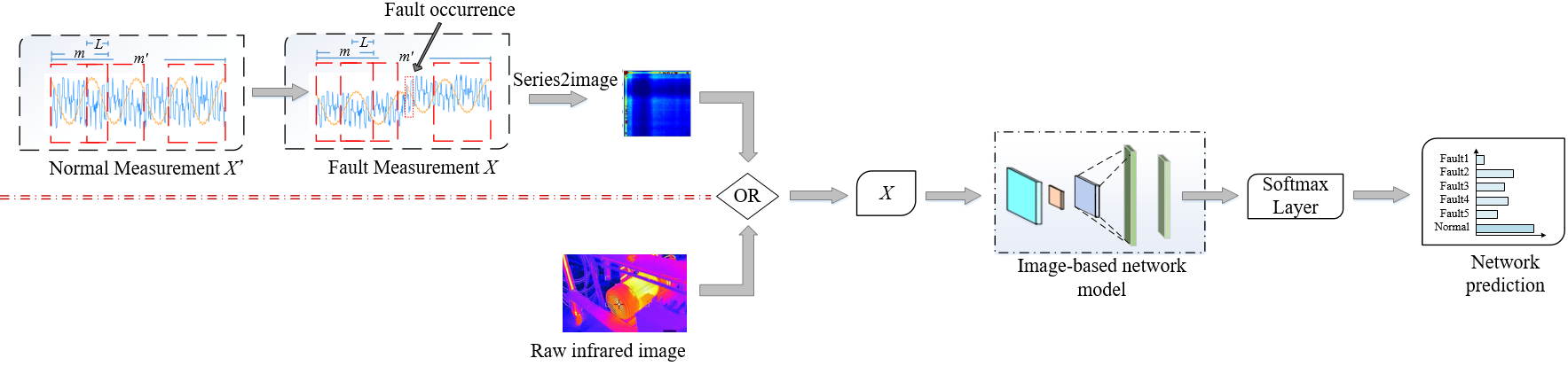}
\caption{Fault diagnosis framework based on image processing. Images are obtained through optical cameras or derived from time-series data \cite{32}. Images go through image processing network, several layers, and a softmax layer, yielding the probability of fault categories.}\label{fig_5}
\end{figure*}

As shown in Fig. \ref{fig_3}, taking a modern wind farm for example, sensor data contains rich heterogeneous information, such as images of possible fan blade cracks and time-series data of vibration, temperature, and speed. After obtaining the sensor data, the NNDF methods can be constructed according to the structures described in Fig. \ref{fig_4} and Fig. \ref{fig_5}, and the typical fault diagnosis is realized.

However, few attention has focused on the ubiquitous dependencies among data. For example, considering the interaction of neighboring wind turbines, a graph representation well models the relationship of all wind turbines in the wind farm, and permits the assessment of the operation status of the whole farm. In fact, even for a single wind turbine, a graph representation models the coupling and dependency among components, and provides additional value for the diagnosis tasks that are hardly achievable when ignoring the functional dependency.

It has been found in \cite{35,36,37} that the category of a given node can be inferred from its neighbors, so except for using time-series and image data for fault diagnosis, the introduction of the graph structure to the data is an alternative and promising way. Furthermore, the involvement of graphs makes it possible for the classical fault diagnosis based on time and space sensing to take into account dependencies type of inductive bias. On top of this, we argue that the connections existing among nodes not only model explicit functional dependencies, but also dependencies as well as topological connections.

Generally speaking, the common methods for processing graph structure among data include the random-walk \cite{Deepwalk}, the graph-clustering \cite{Graphclu}, and GNN-based methods \cite{20}. In this paper, only GNN-based methods, including GCN, graph attention network (GAT), and graph sample and aggregate (GraphSage), are considered.

\section{Graph neural network}

\subsection{Mathematical notations}
Mathematical notations used in this paper follow \cite{52} and are given in Table \ref{tab:nota}. A simple graph can be represented as
\begin{equation}\label{eq_G}
  G=G(V,E)
\end{equation}
where $V$ and $E$ are the sets of nodes and edges, respectively. Let $v_i \in V$ be a node and $e_{ij}=(v_i,v_j)\in E$ denote an edge between $v_i$ and $v_j$. Then, the neighborhood of a node $v$ can be defined as $\mathcal{N}(v)=\{u\in V|(v,u)\in E \}$. Usually, a graph can be described by an adjacency matrix $\mathbf{A} \in \mathcal{R}^{N\times N}$ where $N$ is the number of nodes, that is, $N=|V|$. In particular, $\mathbf{A}_{ij}=1$ if $\{v_i,v_j\}\in E$ and $i\neq j$; $\mathbf{A}_{ij}=1$, otherwise. In undirected graph, $\mathbf{A}_{ij}$ denotes an edge connection between nodes $v_i$ and $v_j$, while in a directed graph, $\mathbf{A}_{ij}$ represents an edge pointing from $v_i$ to $v_j$. In practical applications, a graph may have node features (also called attributes) $\mathbf{X} \in \mathcal{R}^{N\times c}$ where $c$ is the dimension of a node feature vector. A degree matrix $\mathbf{D} \in \mathcal{R}^{N\times N}$ is a diagonal matrix, which can be obtained as $\mathbf{D}_{ii}=\sum_{j=1}^N\mathbf{A}_{ij}$. A graph can also be represented by the Laplacian matrix $\mathbf{L}$, defined as:
\begin{equation}\label{eq_L}
  \mathbf{L}=\mathbf{D}-\mathbf{A}
\end{equation}

\begin{figure*}
\centering
\includegraphics[scale=0.85]{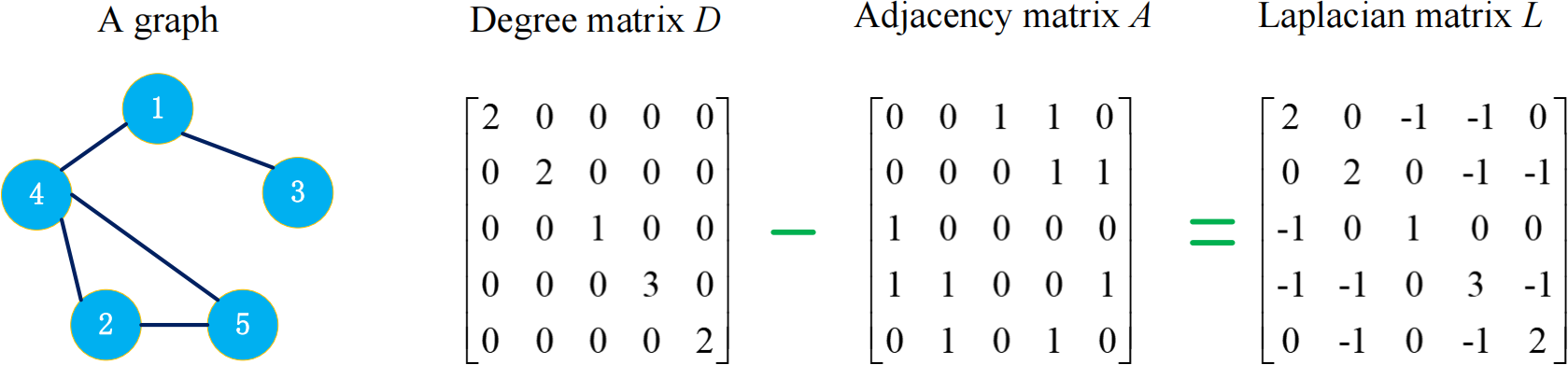}
\caption{Illustration of the Laplacian matrix}\label{fig_ADL}
\end{figure*}

An illustration of the relationship among the above three matrices is shown in Fig. \ref{fig_ADL}.
\begin{table}
  \centering
  \caption{\label{tab:nota} Notations}
    \begin{tabular}{p{43pt} p{185pt}}
        \hline
          \textbf{Notations}    & \textbf{Descriptions}    \\
        \hline
           $||$              & Concatenation operation\\
           $|\cdot|$         & The length of a set \\
           $G$               & A graph representation of a system in the fault-free case \\
           $G_{f}$           & A graph representation of a system in the faulty case \\
           $V$               & The set of nodes in a graph \\
           $v$               & A node $v \in V$ \\
           $E$               & The set of edges in a graph \\
           $\mathcal{N}(v)$  & The neighbors of a node $v$ \\
           $\mathbf{A}$      & The graph adjacency matrix \\
           $\mathbf{A}^T$    & The transpose of the matrix $\mathbf{A}$ \\
           $N$               & The number of nodes, that is, $N=|V|$\\
           $c$               & The dimension of a node feature vector\\
           $\mathbf{X}\in \mathcal{R}^{N\times c}$ & The feature matrix of a graph \\
           $\mathbf{D}$      & The degree matrix of $\mathbf{A}$, that is, $\mathbf{D}_{ii}=\sum_{j=1}^N\mathbf{A}_{ij}$\\
           $\mathbf{L}$      & The Laplacian matrix of a graph, that is, $\mathbf{L}=\mathbf{D}-\mathbf{A}$ \\
           $l$               & The layer index \\
           $t$               & The time step/iteration index\\
           $\delta$          & Nonlinear activation function \\
           $\mathbf{\Theta}$          & Learnable model parameters\\
        \hline
    \end{tabular}
\end{table}

\subsection{GNN Architectures}

The input of GNNs receives two contributions\cite{20}, i.e., the feature matrix $\mathbf{X}$ and the adjacency matrix $\mathbf{A}$. The output of GNNs can be obtained by the general forward propagation equation (\ref{eq_forward})
\begin{equation}\label{eq_forward}
  \mathbf{Z}_{gnn}=softmax(gnn_l(...,gnn_2(\mathbf{A},gnn_1(\mathbf{A},\mathbf{X}))))
\end{equation}
where $gnn$ denotes an operation of GNN and $l$ is the number of layers in the architecture.

\emph{1) Graph convolutional networks (GCNs)}

A GCN can be regarded as an extension of a CNN architecture, and provides an effective method to extract relational/spatial features from graph structured inputs. The GCN model has undergone many improvements, yielding to \cite{43}, which proposed the computation
\begin{equation}\label{eq_output}
\mathbf{Z}=\sigma((\tilde{\mathbf{D}}^{-0.5}\tilde{\mathbf{A}}\tilde{\mathbf{D}}^{-0.5})\mathbf{X}\mathbf{\Theta})
\end{equation}
where $ \tilde{\mathbf{A}}=\mathbf{A}+\mathbf{I}_N$, $\mathbf{I}_N$ is identity matrix of order $N$, and $\tilde{\mathbf{D}}$ a diagonal matrix, whose diagonal elements are $\tilde{\mathbf{D}}_{ii} = \sum_{j=1}^{N}\tilde{\mathbf{A}}_{ij}$. $\sigma$ is an activation function, e.g., the Relu function. $\mathbf{\Theta} \in \mathcal{R}^{c\times c'}$ is the parameter matrix of the network to be learned, $c'$ represents the dimension of the output, and $\mathbf{Z} \in \mathcal{R}^{N \times c'}$ is the output matrix.

The cross-entropy loss function
        \begin{equation}\label{eq_LossF}
        Loss=-\sum_{i=1}^{N_{tr}} \sum_{j=1}^{M} \mathbf{Z}_{ij}log\mathbf{Y}_{ij}
        \end{equation}
is generally used to train the parameters. Here, $N_{tr}$ represents the number of the training points, $M$ is the number of categories, $\mathbf{Z}$ is obtained from Eq. (\ref{eq_output}), and $\mathbf{Y}$ is the corresponding label of the training set. It should be noted that in a classification problem, the dimension of the output equals the number of categories, i.e., $c' = M$.

Even though GCNs have shown major improvements w.r.t. preceding non-relational architectures, there still exists some aspects to be improved:

(1) Shallow GCN network cannot spread label information on a large scale, which limits its receptive field \cite{15}.

(2) The deep GCN network leads to over-smoothing solutions. A multi-layer GCN will make data become highly similar in the forward transmission process, and cannot be correctly distinguished, which greatly reduces the model performance.

(3) GCN belongs to the transductive learning model. When dealing with new data, new nodes must be introduced to modify the original association graph, and the whole GCN model must be trained again to adapt to the new graph.

(4) For a given node, GCN considers all its neighbors to be equally important, instead of paying selective attention to special neighbors: this limits the performance.

Above deficiencies not only restrict performance, but also constraint applications. To address these limits, several new GNN models have been proposed over time.

\emph{2) Graph attention networks}

A GAT architecture uses the attention mechanism to assign different weights to a node neighbors. In addition, it is an inductive model, implying that GAT can perform online (fault diagnosis) task after training \cite{58}\cite{59}.

The GAT architecture is more complex than a GCN one. In detail, the feature matrix $\mathbf{X}=\{\bm{x}_1, \bm{x}_2, \dots ,\bm{x}_N\}\in \mathcal{R}^{N\times c},\bm{x}_i\in \mathcal{R}^c,i=1,2,...,N$. Then, performing a linear transformation on the vector $\bm{x}_i$ of the node to obtain its feature vector $\bm{x}^{'}_i \in \mathcal{R}^{c^{'}}$, $\mathbf{W}$ is a linear mapping matrix:
\begin{eqnarray}
\bm{x}^{'}_i=\mathbf{W}\bm{x}_i, \mathbf{W}\in \mathcal{R}^{c^{'}\times c},i=1,2,...,N
\end{eqnarray}
\begin{eqnarray}
\mathbf{X}^{'}=[\bm{x}^{'}_1, \bm{x}^{'}_2, \dots ,\bm{x}^{'}_N]\in \mathcal{R}^{N\times c^{'}}
\end{eqnarray}

In GAT, for a given $\bm{x}_i$, the attention weight value of its neighbor $\bm{x}_j$ is reflected by $\bm{a}_{ij}$ obtained as $\bm{x}^{'}_i$ and $\bm{x}^{'}_j$ of nodes $i$ and $j$ after linear mapping are concatenated together, and then the inner product is calculated with a vector $\bm{a}$ of length $2c^{'}$. The activation function uses $Leaky\_ReLU$, and a final softmax layer normalizes the weights value.
\begin{eqnarray}
\bm{a}_{ij}=\frac{exp(Leaky\_ReLU([\bm{x}^{'}_i\|\bm{x}^{'}_j]\bm{a}))}{\sum_{k \in \mathcal{N}(i)}exp(Leaky\_ReLU([\bm{x}^{'}_i\|\bm{x}^{'}_k]\bm{a})}
\end{eqnarray}
where $\| $ represents the concatenation operator. Then, without loss of generality, the output of the $(l+1)$th hidden layer become
\begin{equation}\label{eq_gat}
\bm{h}^{l+1}_i=\sigma(\sum_{j\in \mathcal{N}(i)}\mathbf{W}\bm{h}^{l}_j\bm{a}_{ij})
\end{equation}
where $\sigma$ denotes the activation function, and $\bm{h}^{l}$ is the output of the $l$th hidden layer, with $\bm{h}^{0}=\mathbf{X}$. The learnable parameters in GAT include the vector $\bm{a}$ and the parameter matrix $\mathbf{W}$.

\emph{3) Graph sample and aggregate}

Similar to the GAT model, GraphSage belongs to the inductive learning model. In principle, GraphSage learns an aggregation function, which can aggregate the features of a specific node and its neighbors to obtain its high-order features \cite{60}.

Denote the input feature matrix as $\mathbf{X}$. The forward propagation formula of the $l$th hidden layer in a GraphSage model is:
\begin{eqnarray}
\bm{h}^{l}_v=\sigma(\mathbf{W}^l Concat(\bm{h}^{l-1}_v, Aggre({\bm{h}^{l-1}_u, \forall u \in \mathcal{N}(v)}))
\end{eqnarray}
where $Concat$ represents the concatenation operation, $Aggre$ performs an aggregation operation on features, and $\bm{h}^{l}$ the output of the $l$th hidden layer, $\bm{h}^{0}=\mathbf{X}$. Graphsage uses four types of feature aggregation operations, including mean aggregator, GCN aggregator, Long Short-Term Memory (LSTM) aggregator, and a pooling aggregator. GraphSage does not involve the attention mechanism, so it treats all neighbor nodes equally. Due to the inductive learning characteristic, GraphSage can be used for an online fault diagnosis task with a lower computational complexity compared to GCN.

\emph{4) Graph auto-encoder (GAE)}

A GAE is an unsupervised learning model, similar to the auto-encoder \cite{55} \cite{61} \cite{62}. GAE uses a GCN layer as the encoder, its input includes the association graph and the feature matrix, and the output is the coding value $\hat{\mathbf{A}}$ of the node in the graph.
\begin{equation}\label{eq_GAE}
\hat{\mathbf{A}}= \sigma(\mathbf{Z}\mathbf{Z}^T)
\end{equation}
where $\mathbf{Z}=gcn(\mathbf{X},\mathbf{A})$; the mean-squared error function is used as the loss function,
\begin{equation}\label{eq_GAEloss}
L_{GAE} = -\frac{1}{N}\sum_{ij}{||\mathbf{A}_{ij} - \hat{\mathbf{A}}_{ij}||_2^2}
\end{equation}
where $\hat{\mathbf{A}}$ derives from Eq. (\ref{eq_GAE}).

\emph{5) Spatial temporal graph convolutional network (STGCN)}

STGCNs have been widely used to solve traffic forecasting problems \cite{54}. Generally, STGCN is composed both of GCN and CNN architectures, where GCN is responsible for aggregating nodes features in the spatial dimension, while CNN performs temporal convolution in the temporal dimension.

\begin{figure*}
\centering
\includegraphics[scale=0.215]{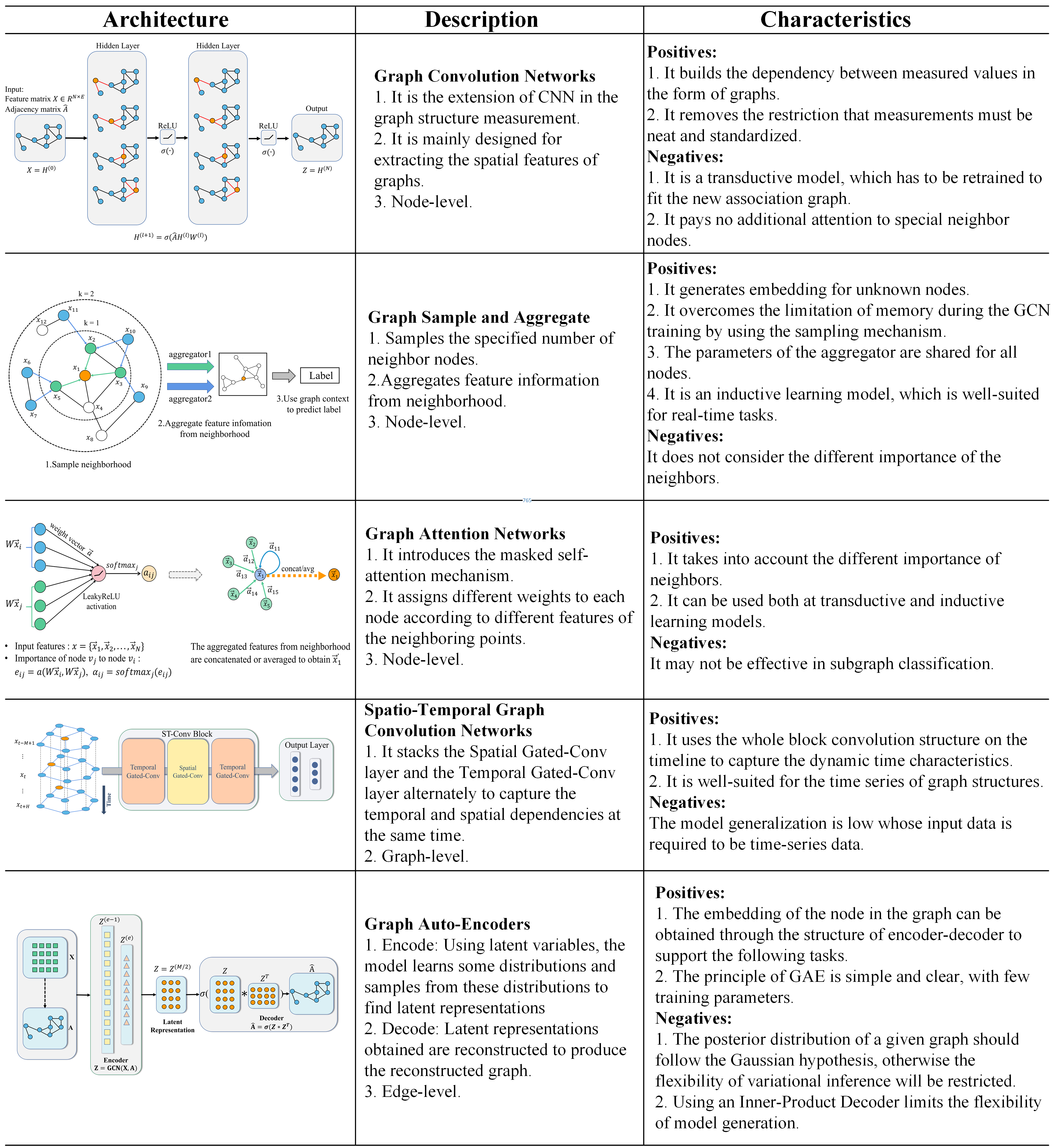}
\caption{A summary of the different GNN architectures}\label{fig_6}
\end{figure*}

According to the different graph analytic tasks, the networks mentioned above fall into three categories, namely, node-level GNNs, edge-level GNNs, and graph-level GNNs \cite{52}. GCN, GAT, and GraphSage belong to the node-level GNN, which classify the nodes in the association graph. Edge-level GNN, like GAE, can be used for matrix completion, which predicts the correlation edges that do not exist in the input adjacency matrix. In a graph-level GNN, such as STGNN, each graph corresponds to a single feature. The characteristics and taxonomy of the aforementioned GNN models in subsection III.B are summarized in Fig. \ref{fig_6}.

\subsection{Visualization of GNN's data classification capability}

\begin{figure}[h!tb]
\begin{center}
\subfigure[Data set without GNN processing]{
\includegraphics[scale=0.388]{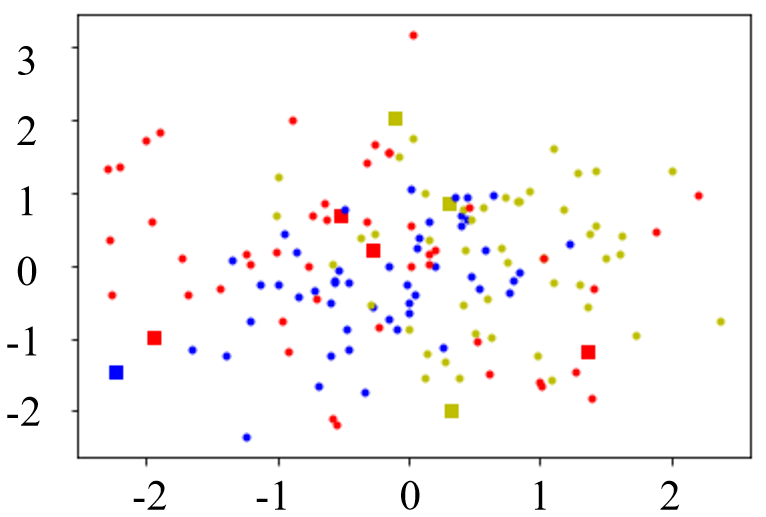}
\label{fig_7a}
}
\\
\subfigure[Data set after a GNN processing]{
\includegraphics[scale=0.388]{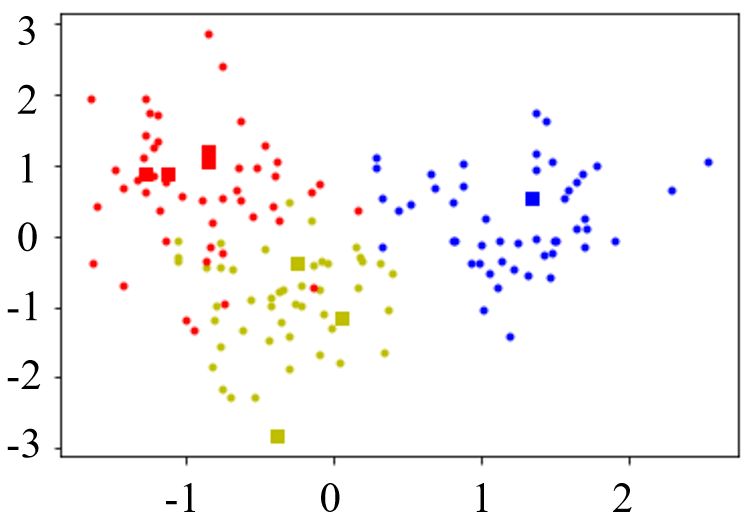}
\label{fig_7b}
}
\caption{Demonstration of GNN's data processing capability (different colors represent different categories of motor running states, square nodes represent the training set, and the circular nodes represent the test set)}
\end{center}
\end{figure}

 Fig. 8 illustrates the efficiency of a graph neural network to exploit information coming from its neighbors. Fig. \ref{fig_7a} shows the original data of a motor running state after normalization and dimensionality reduction. Principal component analysis (PCA) was used to reduce the original data from 48 dimensions to 2 dimensions for visualization. Nodes with different colors show different categories of the motor running states. There are eight square nodes (the number of training sets) in the figure; the test set is composed of 290 nodes. Fig. \ref{fig_7b} shows the motor running state data after the graph convolutional layer. Data have been normalized and their dimension reduced too. The meaning of nodes in both subfigures is the same; data processed by GCN show operational clusters. As shown in Fig. \ref{fig_7a}, it is likely that the high model performance cannot be obtained using the insufficient training set with eight nodes. On the other hand, for the data shown in Fig. \ref{fig_7b}, since nodes belonging to the same categories are more concentrated and nodes belonging to different categories are farther apart, the training set after the GCN layer is more representative. It can be expected that GNN-based method can achieve better classification performance. Considering that the data set for fault diagnosis is always composed of a large proportion of unlabeled data (fault-free data) and a small part of labeled data (faulty data), the ability of GNN to obtain information of neighbors becomes essential.

\section{Fault diagnosis methods based on GNN}

Suppose that an industrial system can be represented by a graph, which consists of nodes and edges. For instance, in a modern chemical process, a great number of sensors are usually installed for indicating the status of the process. Due to the physical coupling, the measurements are entangled or correlated with each other. Hence, if we regard each sensor as a node, then the interactions could be viewed as edges. It should be note that there are various ways to determine nodes and edges. Denote a system operating in a fault-free condition as $G$, and a system in faulty condition as $G_f$. The rationale of a GNN-based fault diagnosis is that
\begin{equation}\label{eq_Gf}
  \{G=(\mathbf{X},\mathbf{A})\}\neq \{G_f=(\mathbf{X}_f,\mathbf{A}_f)\}
\end{equation}
where $\mathbf{X}_f$ and $\mathbf{A}_f$ denote the feature matrix and the adjacency matrix in the faulty case, respectively. Specifically, if a fault occurs in a system, it could affect the features in nodes ($\mathbf{X} \neq \mathbf{X}_f$, $\mathbf{A}=\mathbf{A}_f$), or the adjacency topology ($\mathbf{X} = \mathbf{X}_f$, $\mathbf{A}\neq \mathbf{A}_f$), or both ($\mathbf{X} \neq \mathbf{X}_f$, $\mathbf{A}\neq \mathbf{A}_f$).

In this sense, if the fault information is integrated in $\mathbf{X}$ or $\mathbf{A}$, then the GNN-based fault diagnosis method is feasible. Besides, now that industrial systems are characterized by relational dependencies and that we wish to exploit the relational inductive bias, it is an alternative and promising way to use a GNN-based method for fault diagnosis task.

\begin{figure*}
\centering
\includegraphics[scale=0.45]{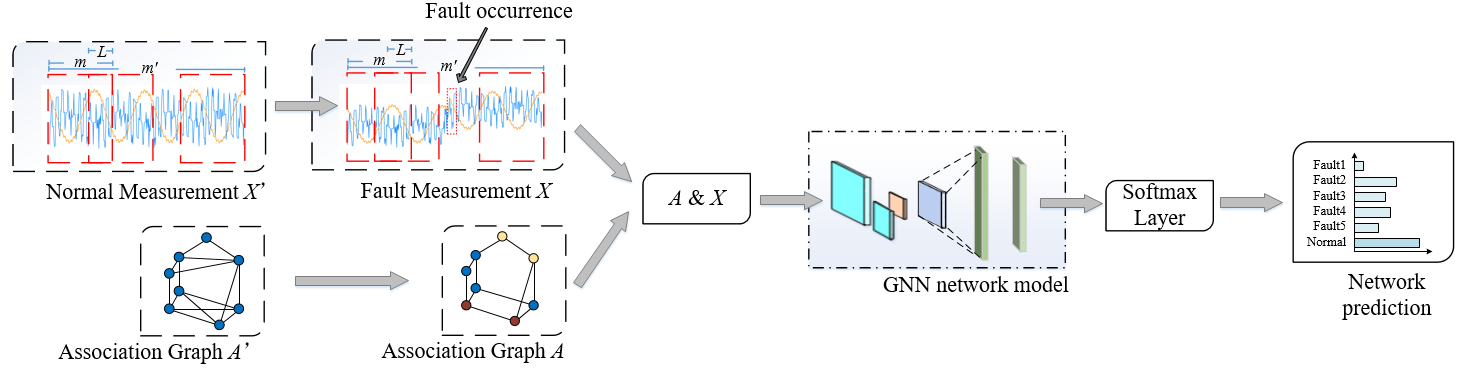}
\caption{Fault diagnosis framework based on GNN}\label{fig_8}
\end{figure*}

Typically, the framework of GNN-based fault diagnosis method is shown in Fig. \ref{fig_8}, which mainly consists of two parts, that is, building the graph from data and building the GNN model.

\subsection{Building association graph from data}

For a given node, a GNN synthetically analyzes the characteristics of a node and its neighbors for fault diagnosis \cite{42}.

According to the difficulty of obtaining the association graph, GNNs can be divided into two categories. In the early GNN-based applications, the association graph of obtained data set has a certain physical meaning, which provides great convenience for its construction and verification. For example, in the Karate Club data set mentioned in \cite{43}, each individual data represents the club member's information, and the task of the neural network is to predict which club a member belongs to. In this example, the association graph is constructed according to club members' interpersonal relationships. The explicit dependency between members simplifies the design of the association graph.

However, in the field of GNN-based fault diagnosis, there is generally no explicit dependency information, so the association graph can only be built from data. How to construct the graph and evaluate the quality of the graph are challenges faced by the GNN-based fault diagnosis method. In this paper, two construction methods for the association graph, as well as their advantages and disadvantages, will be discussed in Section V.

\subsection{Building GNN models for fault diagnosis}
In this subsection, fault diagnosis is transformed into three tasks on graphs, that is node classification, edge classification, and graph classification. In addition, the aforementioned GNN models are integrated into the fault diagnosis algorithms.


\emph{1) Exploration of node-level GNNs for FD}

The considered node-level GNNs in this paper include GCN, GraphSage, and GAT. These architectures treat each measurement as a node in the association graph. As mentioned in subsection III.A.1, two nodes that have a relationship in the association graph are more likely to be divided into the same category by GNN. The implementation of node-level GNN-based FD algorithm is shown in Table \ref{tab:algo1}.

\begin{table}[h]
\centering
\caption{\label{tab:algo1} Algorithm (1)}
 \begin{tabular}[H]{l}
 \hline
 \rule{0pt}{9.5pt}
    Algorithm: Fault\ diagnosis\ method\ based\ on\ node-level\ GNN\\

 \hline
 \ \textbf{Input}: Data set\ $\mathbf{X}$\ with\ length\ $N$, \ adjacency\ matrix $\mathbf{A}$,\ label\ set\ $\mathbf{Y}$\ with\ \\ length\ $N_{tr}$, \ number\ of\ fault\ types $M$,\ number\ of\ iterations $num$.\\

 \rule{0pt}{9.5pt}
 \textbf{Output}: Diagnosis\ results\ of\ the\ test data set\ $\mathbf{X}_{test}$.\\

\rule{0pt}{9.5pt}
1.\  Divide\ data set\ $\mathbf{X}$\ into\ training\ set\ $\mathbf{X}_{train}$, \ test\ set\ $\mathbf{X}_{test}$, \ and \ \\validation\ set\ $\mathbf{X}_{val}$\ according \ to\ the\ label\ set\ $\mathbf{Y}$.\\

\rule{0pt}{9.5pt}
2.\  Construct\ the\ GNN\ model:\\

\rule{0pt}{9.5pt}
3.\ \ \ \ The\ forward\ propagation\ formula\ of\ GNN\ layers\ : \\

\rule{0pt}{9.5pt}
4.\ \ \ \ \ \ \ For\ GCN \ model,\ $\mathbf{Z}=\sigma (\tilde{\mathbf{D}}^{-0.5}\tilde{\mathbf{A}}\tilde{\mathbf{D}}^{-0.5} \mathbf{X}\mathbf{\Theta})$.\\

\rule{0pt}{9.5pt}
5.\ \ \ \ \ \ \ For\ GAT \ model,\ $\mathbf{Z}=\sigma (\sum_{j\in \mathcal{N}(i)}\alpha_{ij}\mathbf{W}\mathbf{X}_j)$.\\

\rule{0pt}{9.5pt}
6.\ \ \ \ \ \ \ For\ GraphSage \ model, \\
\ \ \ \ \ \ \ \ \ \ $\mathbf{Z}=\sigma(\mathbf{W} \cdot Concat(\mathbf{X}_i, Aggregate({\mathbf{X}_j, \forall j \in \mathcal{N}(i)}))$.\\

\rule{0pt}{9.5pt}
7.\ \ \ \ Loss\ function:\ $Loss=-\sum_{i=1}^{N_{tr}} \sum_{j=1}^{M} \mathbf{Z}_{ij}log\mathbf{Y}_{ij}$.\\

\rule{0pt}{9.5pt}
8.\ Train\ the\ GNN\ model: \\

\rule{0pt}{9.5pt}
9.\ \ \ \ for\ $i=1,2,...,num$:\\

\rule{0pt}{9.5pt}
10.\ \ \ \ \ \ \ Input\ the\ total\ data set\ $\mathbf{X}$\ to\ the\ GNN\ model.\\

\rule{0pt}{9.5pt}
11.\ \ \ \ \ \ \ Calculate\ the\ loss\ function.\\

\rule{0pt}{9.5pt}
12.\ \ \ \ \ \ \ Update\ the\ model\ with\ back\ propagation.\\

\rule{0pt}{9.5pt}
13.\ \ \ \ Complete\ GNN\ model\ training.\\

\rule{0pt}{9.5pt}
14. Validate\ the\ trained\ GNN\ model\ using\ the\ data set\ $\mathbf{X}_{val}$.\\

\rule{0pt}{9.5pt}
15. Obtain\ the\ diagnosis\ results, \ $\mathbf{Z}=model(\mathbf{X}_{test})$.\\

    \hline
 \end{tabular}
\end{table}

\emph{2) Exploration of edge-level GNNs in fault diagnosis}

In this paper, edge-level GNN is chosen as a GAE. As its outputs are the reconstruction of the corresponding adjacency matrices of the graph, it cannot be directly used for fault diagnosis. However, it is pointed out in subsection V.A that a ready-made association graph in fault diagnosis is often not accessible, and the accuracy of the graph constructed by various methods cannot be guaranteed. Therefore, the edge-level GNN is used to reconstruct the original association graph, so that nodes with uncertain relationship (the corresponding value in the adjacency matrix is 0) are identified as having association. The above steps construct the association graph that can better reflect dependencies between nodes.

Edge-level GNN can be combined with node-level GNN. First, the original graph is input into the GAE model for reconstruction, then the reconstructed graph is fed into the node-level GNN together with feature matrix $\mathbf{X}$ to realize the final fault diagnosis. The implementation of the edge-level GNN-based FD algorithm is given in Table \ref{tab:algo2} (only GAE model is considered in this paper).

\begin{table}[h]
\centering
\caption{\label{tab:algo2} Algorithm (2)}
 \begin{tabular}[H]{l}
 \hline
 \rule{0pt}{10pt}
    Algorithm: Fault\ diagnosis\ method\ based\ on\ edge-level\ GNN\\

 \hline
 \ \textbf{Input}: Data set\ $\mathbf{X}$\ with\ length\ $N$, \ adjacency\ matrix $\mathbf{A}$,\ number\ of\ \\ iterations $num$.\\

 \rule{0pt}{10pt}
 \textbf{Output}: The\ reconstruction\ matrix\ $\hat{\mathbf{A}}$\ of\ the\ adjacency\ matrix\ $\mathbf{A}$.\\

\rule{0pt}{10pt}
1.\  Construct\ the\ GAE\ model:\\

\rule{0pt}{10pt}
2.\ \ \ \ The\ forward\ propagation\ formula\ of\ GAE\ layers\ : \\

\rule{0pt}{10pt}
3.\ \ \ \ \ \ \ $\mathbf{Z}=gcn(\mathbf{X}, \mathbf{A})$.\\

\rule{0pt}{10pt}
4.\ \ \ \ \ \ \ $\hat{\mathbf{A}}=\sigma(\mathbf{Z}\mathbf{Z}^T)$.\\

\rule{0pt}{10pt}
5.\ \ \ \ Loss\ function:\ \\
 \ \ \ \ \ \ \ $Loss=-\frac{1}{N}\sum \mathbf{A}_{ij}log\hat{\mathbf{A}}_{ij}+(1-\mathbf{A}_{ij})log(1-\hat{\mathbf{A}}_{ij})$.\\

\rule{0pt}{10pt}
6.\ Train\ the\ GAE\ model: \\

\rule{0pt}{10pt}
7.\ \ \ \ for\ $i=1,2,...,num$:\\

\rule{0pt}{10pt}
8.\ \ \ \ \ \ \ Input\ the\ data set\ $\mathbf{X}$\ and\ the\ matrix\ $\mathbf{A}$\ to\ the\ GAE\ model. \\

\rule{0pt}{10pt}
9.\ \ \ \ \ \ \ Calculate\ the\ loss\ function.\\

\rule{0pt}{10pt}
10.\ \ \ \ \ \ \ Update\ the\ model\ with\ back\ propagation.\\

\rule{0pt}{10pt}
11. Obtain\ the\ reconstruction\ matrix\ $\hat{\mathbf{A}}$.\\

    \hline
 \end{tabular}
\end{table}

\emph{3) Exploration of graph-level GNNs in fault diagnosis}

STGCN is the commonly used graph-level GNN. Unlike node-level GNN, STGCN does not focus on the dependencies within the data set, but on the complex dependency between components in a single data point \cite{54}. In the task of fault diagnosis, time-series are usually composed of data collected by multiple sensors. Therefore, a sample can be understood as a node, and a series of measurements sampled by multiple sensors together form a graph. The implementation of the graph-level GNN-based FD algorithm is shown in Table \ref{tab:algo3}.

%

\begin{table}[h]
\centering
\caption{\label{tab:algo3} Algorithm (3)}
 \begin{tabular}[H]{l}
 \hline
 \rule{0pt}{10pt}
    Algorithm: Fault\ diagnosis\ method\ based\ on\ graph-level\ GNN\\

 \hline
 \ \textbf{Input}: Data set\ $\mathbf{X}$\ with\ length\ $N$, \ adjacency\ matrix $\mathbf{A}$,\ label\ set\ $\mathbf{Y}_L$\ with\ \\ length\ $n_1$, number of fault\ types $M$,\ number\ of\ iterations $num$.\\

 \rule{0pt}{10pt}
 \textbf{Output}: Diagnosis\ results\ of\ the test\ data set\ $\mathbf{X}_{test}$.\\

\rule{0pt}{10pt}
1.\ Divide\ dataset\ $\mathbf{X}$\ into\ training\ set\ $\mathbf{X}_{train}$,\ test\ set\  $\mathbf{X}_{test}$\ and\ \\ validation set\ $\mathbf{X}_{val}$ according\ to\ the\ label\ set\ $\mathbf{Y}_L$.\\

\rule{0pt}{10pt}
2.\ Use\ $\mathbf{X}_{test}$\ to\ construct\ the\ adjacency\ matrix\ $\mathbf{X}$.\\

\rule{0pt}{10pt}
3.\ Construct\ the\ STGCN\ model:\\

\rule{0pt}{10pt}
4.\ \ \ \ In\ STGCN\ Block\ : \\

\rule{0pt}{10pt}
5.\ \ \ \ \ \ \ \ In\ Temporal\ Block1 \ : \\

\rule{0pt}{10pt}
6.\ \ \ \ \ \ \ \ \ \ \ $\mathbf{X}_{out1} = CNNs(\mathbf{X})$.\\

\rule{0pt}{10pt}
7.\ \ \ \ \ \ \ \ In\ Spatial\ Block \ : \\

\rule{0pt}{10pt}
8.\ \ \ \ \ \ \ \ \ \ \ $\mathbf{X}_{out2}=\sigma (\tilde{\mathbf{D}}^{-0.5}\tilde{\mathbf{A}}\tilde{\mathbf{D}}^{-0.5} \mathbf{X}_{out1}\mathbf{\Theta})$.\\

\rule{0pt}{10pt}
9.\ \ \ \ \ \ \ \ In\ Temporal\ Block2 \ : \\

\rule{0pt}{10pt}
10.\ \ \ \ \ \ \ \ \ \ \ $\mathbf{Z} = CNNs(\mathbf{X}_{out2})$.\\

\rule{0pt}{10pt}
11.\ \ \ \ Loss\ function:\ $Loss=-\sum_{l=1}^{n_1} \sum_{f=1}^{M} \mathbf{Z}_{lf}logL_{lf}$.\\

\rule{0pt}{10pt}
12.\ Train\ the\ STGCN\ model: \\

\rule{0pt}{10pt}
13.\ \ \ \ for\ $i=1,2,...,num$:\\

\rule{0pt}{10pt}
14.\ \ \ \ \ \ \ Input\ the\ training\ set\ $\mathbf{X}_{train}$ \ to\ the\ STGCN\ model.\\

\rule{0pt}{10pt}
15.\ \ \ \ \ \ \ Calculate\ the\ loss\ function.\\

\rule{0pt}{10pt}
16.\ \ \ \ \ \ \ Update\ the\ model\ with\ back\ propagation.\\

\rule{0pt}{10pt}
17.\ \ \ \ \ \ \ Evaluate\ the\ model\ with\ validation\ set\ $\mathbf{X}_{val}$.\\

\rule{0pt}{10pt}
18.\ \ \ \ Complete\ GNN\ model\ training.\\

\rule{0pt}{10pt}
19. Validate\ the\ trained\ STGNN\ model\ using\ the\ data set\ $\mathbf{X}_{val}$.\\

\rule{0pt}{10pt}
20. Obtain\ the\ diagnosis\ results, \ $\mathbf{Z}=model(\mathbf{X}_{test})$.\\

    \hline
 \end{tabular}
\end{table}

\section{Discussions on constructing association graph}

As mentioned previously, the association graph is an essential part of GNN, which reflects the implicit dependency among nodes. At present, the widely used construction methods for the association graph are:

\subsection{Using K-nearest neighbor method to construct the association graph}

In practice, a possible method to construct the association graph is: first, a large number of time-frequency features of a given data set are extracted. Then, a small number of time-frequency features that can reflect the characteristic of the system are obtained through feature selection. Next the K-nearest neighbor (KNN) method is used to find the $K$ nearest neighbors of the given node according to these features. Finally, the construction of the association graph is realized \cite{45}.

Suppose the data set $\mathbf{X}=\{\bm{x}_1,\bm{x}_2,...,\bm{x}_N\}$ is a feature matrix, the KNN method is used, then the nearest neighbor matrix $\mathbf{P}$ is formed, where $\mathbf{P}_i$ stores the $K$ nearest neighbor nodes of the $i$th data in $\mathbf{X}$, $i\in [1,N]$. The composition can be realized by using the nearest neighbor matrix $\mathbf{P}$, and the corresponding adjacency matrix $\mathbf{A} \in \mathcal{R}^{N\times N}$, that is,
\begin{equation}
\mathbf{A}_{ij}=\left\{
    \begin{array}{rcl}
    1   &    & \mathbf{P}_i \in \mathbf{x}_j                      \\
    0   &    & \mathbf{P}_i \notin \mathbf{x}_j                   \\
    \end{array} \right.
\end{equation}

The illustration of KNN-based graph construction method is shown in Fig. \ref{fig_KNN}, and there are two factors that affect the quality of KNN-based graph construction method:

\begin{figure}
\centering
\includegraphics[scale=0.51]{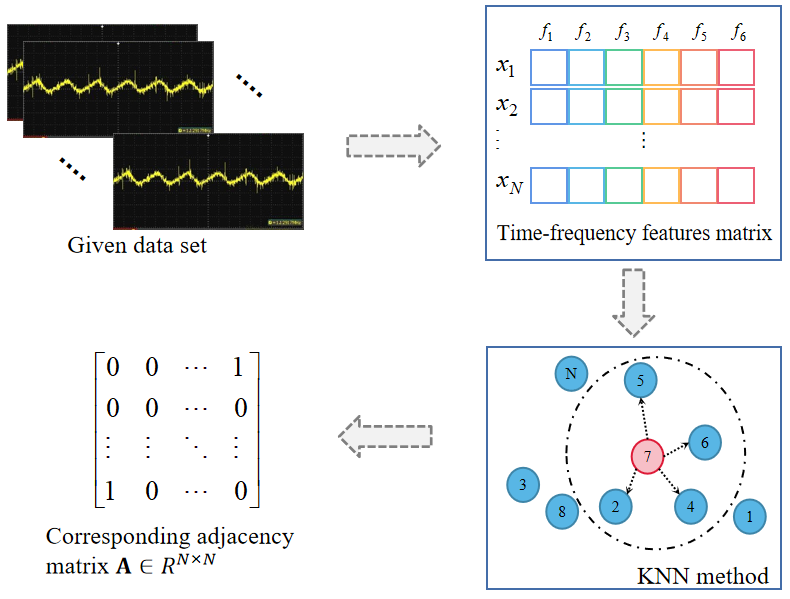}
\caption{KNN-based graph construction method}\label{fig_KNN}
\end{figure}

\begin{enumerate}
  \item The selection of time and frequency domain features \cite{46}. In fact, the selection of feature types is completely subjective. It is difficult to explain why some feature types are used instead of others, many trial-and-error experiments are necessary.
  \item The determination of the distance between nodes. A straightforward method is to use Euclidean distance to measure the distance between features. From another point of view, the occurrence of a fault will lead to the deviation of features, but may not affect others. Therefore, various features should be lead to different distance values. However, due to the principle of KNN method, all features are treated equally. In addition, its time complexity is $O(N\times N)$, resulting that the increase of the amount of data $N$ can lead to a dramatic increase of the time consumption for the graph construction.
\end{enumerate}

In general, the KNN-based construction method is favorable due to its simple and clear logic.


\subsection{Using prior knowledge to construct the association graph}

\begin{figure}
\centering
\includegraphics[scale=0.42]{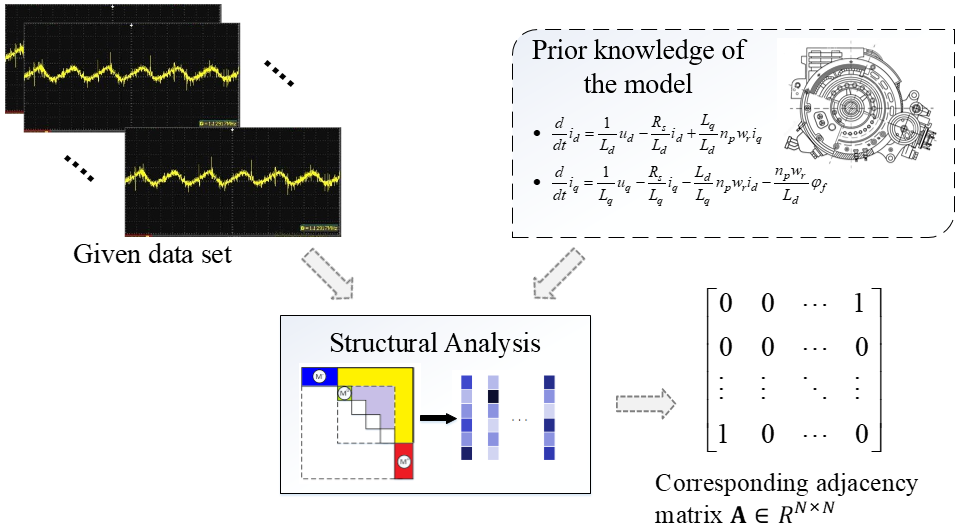}
\caption{SA-based graph construction method}\label{fig_SA}
\end{figure}

In \cite{47}, the idea of using prior knowledge to construct the association graph is considered. It uses structural analysis (SA) method to prediagnose the fault, and transforms the results of prediagnosis into a graph to construct the GCN model for the final fault diagnosis. In this method, the graph construction is a bridge that connects the model-based SA and the data-based GCN. The illustration of SA-based graph construction method is shown in Fig. \ref{fig_SA}.

The introduction of prior knowledge significantly increases the training time, and requires not only the knowledge about the mechanisms of the system, but also data information. On the other hand, it ensures the accuracy of the graph, greatly improves the overall performance of the model, and is one of the best methods to build the association graph when prior knowledge is available. In addition, it should be mentioned that, as a bridge, the association graph can combine the knowledge with the measurements to exploit the strengths of both methods.

\subsection{Using matrix completion to adjust the association graph}

The construction process of the graph is to analyze the potential existence of the dependency between nodes. Therefore, the construction of the association graph can also be transformed to the link prediction issue. Inspired by \cite{48}, we propose a new method for constructing the association graph. Firstly, an incomplete association graph of a data set is constructed by the KNN method. Then the adjacency matrix of the graph is reconstructed by the GAE model. Finally, the downstream task is completed according to the reconstructed association graph.

\begin{figure*}
\centering
\includegraphics[scale=0.58]{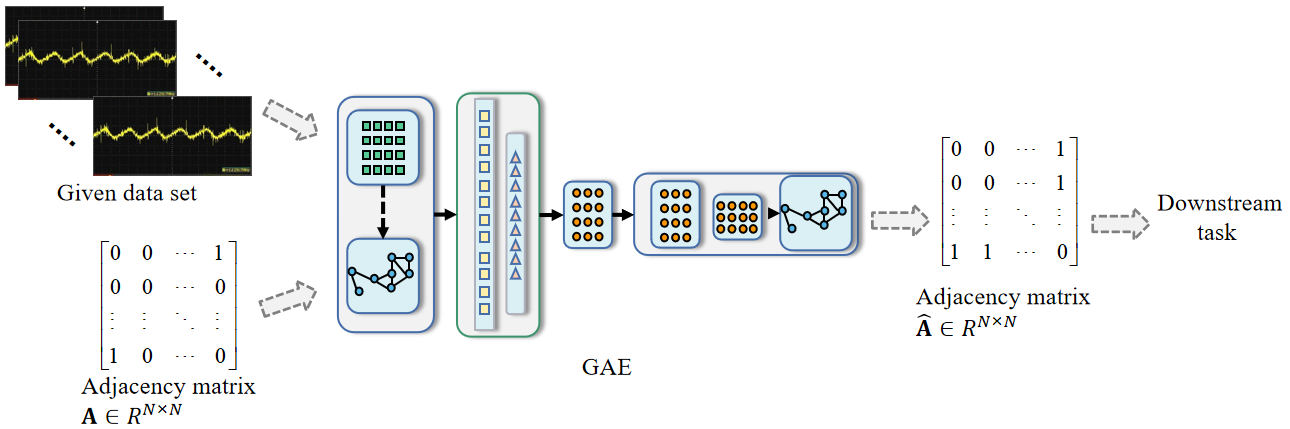}
\caption{GAE-based graph construction method}\label{fig_GAE}
\end{figure*}

The method of using a matrix to complete and adjust the association graph is an extension of the KNN method. Since GAE is a type of unsupervised learning, its purpose is to adjust the original association graph, so the graph reconstructed by GAE is at least no worse than the original one. Subsequent experiments show that the method of reconstructing association graph using GAE model can improve diagnostic performance to a certain extent. The illustration of GAE-based graph adjustment method is shown in Fig. \ref{fig_GAE}.

\subsection{Assess the quality of the association graph}

According to the constructed association graphs, GNNs aggregate the isolated data into a whole. Through various aggregation methods, GNNs enable each node to aggregate the information of its neighbors. Due to this, the performance of the GNN model is better than that of ordinary neural network in many fields.

In \cite{49}, it was pointed out that the GNN model has been widely used in graph representation learning, but it remains a issue to measure the quality of the graph. To deal with this, two quality indicators, namely feature smoothness and label smoothness, were proposed. The feature smoothness index $\lambda_f$ is:
\begin{equation}
\lambda_f=\frac{||\sum_{e_{i,j}\in E}(\bm{x}_{v_i}-\bm{x}_{v_j})^2||}{|E|\cdot d}
\end{equation}
where $\bm{x}_{v_i}$ and $\bm{x}_{v_j}$ represent the features of the nodes $v_i$ and $v_j$, respectively, $d$ is the dimension of $\bm{x}_{v_i}$ and $\bm{x}_{v_j}$, and $|E|$ represents the amount of the edges $E$ in a association graph. It is assumed that nodes with dissimilar features compared with their neighbors tend to obtain more information from association graph. Therefore, $\lambda_f$ is positively correlated with the quality of the association graph.

Another label smoothness index $\lambda_l$ is defined as:
\begin{equation}
\lambda_l=\sum_{e_{i,j}\in E}(1-\mathbb{I}(v_i, v_j))/|E|
\end{equation}
where $\mathbb{I}(v_i, v_j)=0$ if $Y_{v_i}\neq Y_{v_j}$, $\mathbb{I}(v_i, v_j)=1$ if $Y_{v_i}=Y_{v_j}$, $Y_{v_i}$ and $Y_{v_j}$ represent the category of the nodes $v_i$ and $v_j$, respectively.

The purpose of calculating $\lambda_l$ is based on an inductive bias that, in a graph, if most nodes and their neighbors are in the same categories, the graph is beneficial for the training of the model. Therefore, $\lambda_l$ is negatively correlated with the quality of the association graph.

In conclusion, $\lambda_f$ and $\lambda_l$ measure the quality of the association graph from two aspects. They point out that a high-quality graph should have the following characteristics: the feature distributions of two neighbors in the graph are quite different (corresponding to a big $\lambda_f$ indicator), but they have the same category (corresponding to a small $\lambda_l$ indicator).

Interestingly, \cite{49} implicitly explains the limitation of KNN-based construction method: it tends to connect two nodes with similar feature distribution, which makes nodes obtain less information from their neighbors. From this point of view, both the $\lambda_f$ index and the $\lambda_l$ index calculated by the graph from KNN-based method are small, which indicate a ``correct but useless'' association graph.

\section{Benchmark study and comparison}

To illustrate the performance of GNN-based FD methods, three baseline models as well as several GNN-based FD methods are tested on three industrial data sets and compared.

\subsection{The designed models}

The detailed architectures of various GNN methods have been discussed in Section III. The corresponding hyperparameters of each models are shown in Table \ref{tab:Hyper}. The GCN model is composed of a GC layer, three 1dCNN layers, and two fully connected layers; The GAT model is composed of two graph attention layers with a multihead mechanism, where the number of heads is eight; The GraphSage model is composed of two layers, and the GCN aggregator is used; the STGCN model contains two temporal blocks and one spatial block. The spatial block uses traditional GCN for graph convolution, while the temporal blocks perform convolution-max-pooling-convolution for each node in the temporal dimension.

\begin{table}
  \centering
  \caption{\label{tab:Hyper} Hyperparametric numerical simulation of GNN model}
    \begin{tabular}[H]{ c  c  c  c   }
        \hline
        \rule{0pt}{8pt}
          Models       & Number of epochs & Learning rate & Optimizer    \\
        \hline
        \rule{0pt}{8pt}
           GCN\        & 300 & 0.000 18 & RMSProp \\
           GAT\        & 200 & 0.000 05 & Adam    \\
           GraphSage\  & 300 & 0.005   & RMSProp \\
           STGCN\      & 200 & 0.001   & Adam    \\
        \hline
    \end{tabular}
\end{table}

Furthermore, 6 kinds of widely-used baseline models are used, namely the CNN, LSTM, RF, GBDT, LGBM, and SVC models.

1. \textbf{CNN}. CNN is a rather common baseline model, which performs well in dealing with issues in various research fields.

2. \textbf{LSTM}. LSTM does well in processing time-series data. It is kind of an advanced recurrent neural network (RNN).

3. \textbf{RF}. Random forest model (RF) constructs multiple decision trees. To realize the prediction purpose, it counts the prediction results of each decision tree, and obtains the final results through voting methods.

4. \textbf{GBDT}. Gradient boosting decision tree, also known as MART (multiple additive regression tree), is an iterative decision tree method. The algorithm is composed of multiple decision trees, and the conclusions of all trees are accumulated to make the final answer.

5. \textbf{LGBM}. Light gradient boosting machine is a framework for implementing GBDT method. It supports efficient parallel training, and has the advantages of faster training speed, lower memory consumption, better accuracy, distributed support and rapid processing of massive data.

6. \textbf{SVC}. Support vector classification avoids the traditional process from induction to deduction, efficiently realizes the transformation from training process to prediction process, and greatly simplifies the usual problems such as classification and regression.

\subsection{Description of data sets}

\begin{figure}
\centering
\includegraphics[scale=0.34]{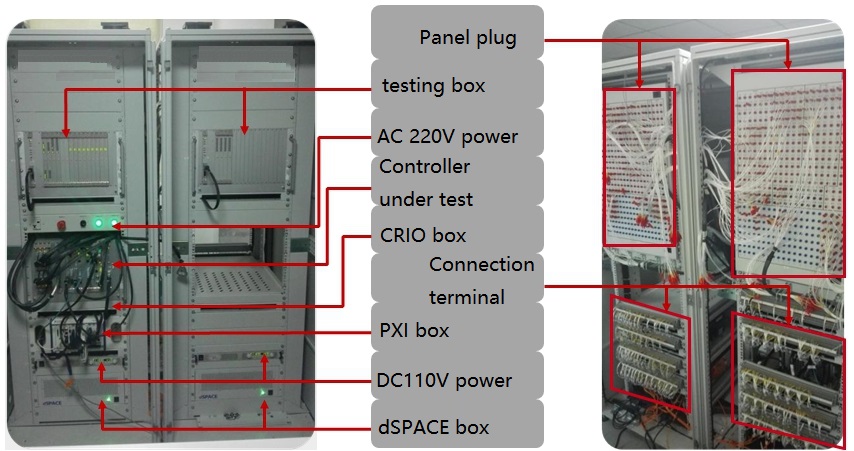}
\caption{Hardware-in-the-loop simulation platform for the pulse rectifier of a traction system}\label{fig_9}
\end{figure}

The experiments in this paper are implemented on three data sets. The first data set is obtained from a hardware-in-the-loop (HIL) simulation platform based on the structure of the pulse rectifier in a traction system (as shown in Fig. \ref{fig_9}), named the ``rectifier data set''. The second data set is collected from a motor benchmark, named ``motor data set''. The last data set is obtained from the Tennessee Eastman chemical process benchmark, named the ``TEP data set''\footnote{\text{http://brahms.scs.uiuc.edu}}.

\begin{table}
  \centering
  \caption{Description of faults}
  \label{tab:Ftype}\center
    \begin{tabular}{m{30pt} m{120pt} m{60pt}} 
        \hline
     {Data set}   & {Description of fault}    & {Fault type}  \\ \hline
     ~           & Current sensor fault & Random variation \\
     ~           & IGBT1 fault  & Step \\
     Rectifier   & IGBT2 fault  & Step \\
     ~           & IGBT3 fault  & Step \\
     ~           & IGBT4 fault  & Step \\
     ~           & Demagnetization fault    & Slow drifting \\
     Motor       & Inter-turn short fault   & Step \\
     ~           & Bearing fault            & Random variation \\
     ~           & A/C feed ratio, B composition constant   & Step \\
     ~           & B composition, A/C ratio constant        & Step \\
     ~           & D feed temperature                       & Step \\
     TEP         & Reactor cooling water inlet temp.        & Step \\
     ~           & Condenser cooling water inlet temp.      & Step \\
     ~           & A feed loss                              & Step \\
     ~           & C Header pressure loss - reduced availability   & Step \\
        \hline
    \end{tabular}
\end{table}

The dimension of the rectifier data set is 1067 $\times$ 256 $\times$ 6. It is composed of 1067 labeled measurements. Each measurement is composed of six sensor samples, and the dimension of a single measurement is 256 $\times$ 6. The 1067 measurements are divided into one normal condition and five fault types.

The motor data set consists of 1104 labeled measurements, each of which records 48 features of its corresponding motor, and the dimension of the total data set is 1104 $\times$ 48. It contains one normal condition and three fault types. Since the motor data set has no evident sequential characteristic, STGCN is not used for validation in it.

The dimension of TEP data set is 1100 $\times$ 500 $\times$ 52, which is composed of 1100 labeled measurements, each of which is composed of 41 observation variables and 11 control variables. It contains one normal condition and seven fault types. Fault types exist in the three data sets are briefly described in Table \ref{tab:Ftype}.

The three data sets are collected from systems either with known structural information (like the pulse rectifier) or without known structural information. Thereby, they are representative in the field of fault diagnosis. Related parameters of the data sets are summarized in Table \ref{tab:Dataset}.

\begin{table*}
  \centering
  \caption{\label{tab:Dataset} Details of three data sets}
    \begin{tabular}[H]{ c  c  c  c  c  }
        \hline
        \rule{0pt}{8pt}
         Data set     &   Quantity       & Dimensions    &  Structural information &   Data complexity     \\
        \hline
        \rule{0pt}{8pt}
           rectifier data set        &   1067  &  256$\times 6$            &  Available     &  High   \\
           motor data set            &   1104  &       48                  &  Unavailable   &  Low    \\
           TEP data set              &   1100  &  5000$\times52$           &  Unavailable   &  High   \\

        \hline
    \end{tabular}
\end{table*}

\subsection{Description of fault types}

As is shown in Table \ref{tab:Ftype}, more than ten types of faults are considered in experiments, which can be roughly divided into three categories, including random variation fault, step fault, and slow drifting fault. Their characteristics are depicted as follows \cite{YC}.

1. \textbf{Characteristics of a random variation fault}. Sampled values with random variation fault will fluctuate randomly. Therefore, there exists irregularly changed deviations between the samples and the actual value. The generation of the random variation faults is affected by many factors including noise and uncertainties, which are ubiquitous in a industrial process.

2. \textbf{Characteristics of a step fault}. The deviation between sampled values and actual values changes significantly in a short period of time. The step fault can lead to great harm to a industrial system, take pulse rectifier for example, step fault may be caused by an open circuit or short circuit.

3. \textbf{Characteristics of a slow drifting fault}. The samples adds an extra signal that is proportional to time. Slow drifting faults may result from aging of components.

\subsection{Constructing association graphs from the data sets}

The construction of the association graph plays an important role in GNN-based fault diagnosis. Based on SA, KNN, and KNN + GAE methods, three kinds of graphs are constructed.

\begin{figure}
\begin{center}
\subfigure[Graph constructed by SA method]{
\includegraphics[width=0.27\textwidth]{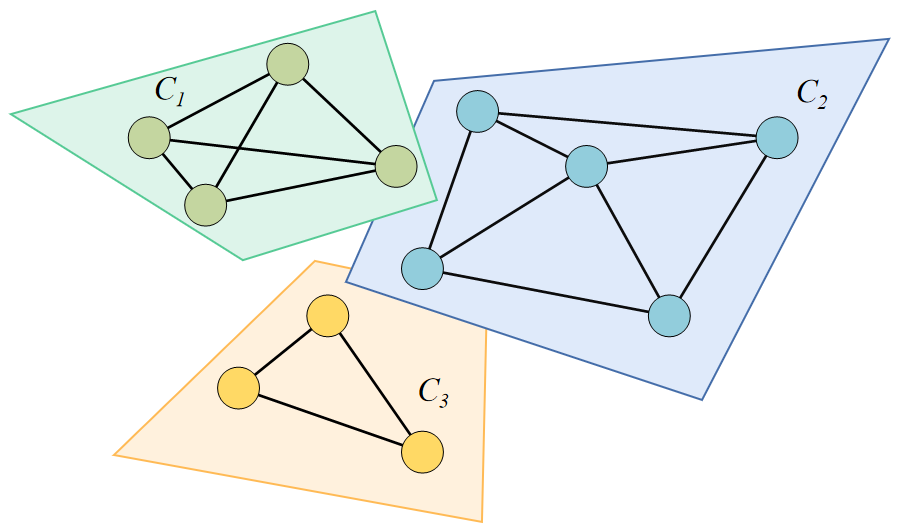}
\label{fig:SA-Graph}
}
\subfigure[Graph constructed by KNN method]{
\includegraphics[width=0.16\textwidth]{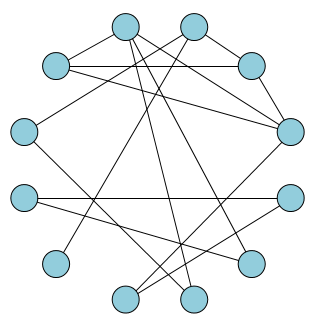}
\label{fig:KNN-Graph}
}
\\
\subfigure[Graph constructed by KNN+GAE method]{
\includegraphics[width=0.16\textwidth]{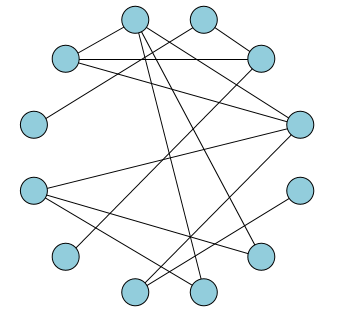}
\label{fig:KNN-GAE-Graph}
}
\caption{Schematic diagrams of graphs constructed by three kinds of methods}\label{fig_graph}
\end{center}
\end{figure}

Among graphs constructed by three kinds of methods, each of them has its own characteristics. For the graph constructed by SA method, all nodes in a graph are clustered respectively. Nodes belonging to the same cluster are connected to each other, and nodes of different clusters are isolated from each other (as is shown in Fig. \ref{fig:SA-Graph}). For the graph constructed by KNN method, all nodes have the potential to be connected to each other. Usually, all nodes in the graph form a connected domain (as is shown in Fig. \ref{fig:KNN-Graph}). The graph constructed by KNN + GAE method is slightly different from the KNN-based graph, whose schematic diagram is shown in Fig. \ref{fig:KNN-GAE-Graph}.

Related parameters about each graph are shown in Table \ref{tab:Graph}. Note that the $K$ value is only applicable to the KNN method.

\begin{table}[!t]
  \centering
  \caption{\label{tab:Graph} Related parameters of the association graph}
    \begin{tabular}[H]{ c  c  c  c }
        \hline
        Construction method & Data set        &  No. of edges  & $K$ value        \\
        \hline
            SA        &   rectifier data set  &   104 521     &  $\backslash$       \\
                      &   rectifier data set  &   325 56      &  45                 \\
            KNN       &   motor data set      &   540 96      &  50                 \\
                      &   TEP data set        &   539 00      &  30                 \\
                      &   rectifier data set  &   862 69      &  45                 \\
            KNN+GAE   &   motor data set      &   119 387     &  50                 \\
                      &   TEP data set        &   103 667     &  30                 \\
        \hline
    \end{tabular}
\end{table}

\subsection{Experimental results and analysis}

\subsubsection{Comparison between graph construction methods}

The sizes of training set in three data sets are

The amount of training set in the rectifier data set is 50 (accounts for 4.69$\%$ of the data set), while 80 in the motor data set (accounts for 7.25$\%$ of the data set), and 150 in the TEP data set (accounts for 13.64$\%$ of the data set). To make a comparison between different construction methods, we used the SA-based method, the KNN and KNN + GAE-based method on the rectifier data set, KNN and KNN + GAE-based methods on the motor and TEP data sets. The diagnostic accuracies are shown in Table \ref{tab:Result}, they reflect the performance of various diagnostic methods. For example, when the size of the test set is 867, and the accurate prediction quantity is 617, the diagnostic accuracy is calculated as 0.712.

\begin{table*}[]
\center
\caption{\label{tab:Result} Experimental results (accuracy$\pm$standard deviation)}
\begin{tabular}{cccccc}
\hline
Model                            & \multicolumn{2}{c}{Method}           & rectifier data set(Train\_set = 50) & motor data set(Train\_set = 80) & TEP data set(Train\_set = 150) \\ \hline
\multirow{10}{*}{GNN-Models}     & \multirow{3}{*}{GCN}       & SA      & 0.862$\pm$0.042                     & \textbackslash{}                & \textbackslash{}               \\
                                 &                            & KNN     & 0.712$\pm$0.028                     & 0.747$\pm$0.018                 & 0.993$\pm$0.002                          \\
                                 &                            & KNN+GAE & 0.750$\pm$0.011                     & 0.736$\pm$0.030                 & 0.970$\pm$0.051                          \\ \cline{2-6}
                                 & \multirow{3}{*}{GAT}       & SA      & 0.901$\pm$0.056                     & \textbackslash{}                & \textbackslash{}               \\
                                 &                            & KNN     & 0.801$\pm$0.024                     & 0.661$\pm$0.036                 & 0.969$\pm$0.033                          \\
                                 &                            & KNN+GAE & 0.877$\pm$0.017                     & 0.697$\pm$0.041                 & 0.988$\pm$0.010                          \\ \cline{2-6}
                                 & \multirow{3}{*}{Graphsage} & SA      & 0.943$\pm$0.001                     & \textbackslash{}                & \textbackslash{}               \\
                                 &                            & KNN     & 0.786$\pm$0.011                     & 0.809$\pm$0.014                 & 0.993$\pm$0.002                          \\
                                 &                            & KNN+GAE & 0.818$\pm$0.006                     & 0.823$\pm$0.014                 & 0.996$\pm$0.001                          \\ \cline{2-6}
                                 & \multicolumn{2}{c}{STGCN}            & 0.843$\pm$0.141                     & \textbackslash{}                & 0.976$\pm$0.007                          \\ \hline
\multirow{6}{*}{Baseline-Models} & \multicolumn{2}{c}{CNN}              & 0.842$\pm$0.047                     & 0.472$\pm$0.079                 & 0.988$\pm$0.013                          \\
                                 & \multicolumn{2}{c}{LSTM}             & 0.675$\pm$0.046                     & 0.442$\pm$0.075                 & 0.912$\pm$0.016                          \\
                                 & \multicolumn{2}{c}{RF}               & 0.666$\pm$0.015                     & 0.547$\pm$0.067                 & 0.855$\pm$0.005                          \\
                                 & \multicolumn{2}{c}{LGBM}             & 0.457$\pm$0.016                     & 0.958$\pm$0.020                 & 0.938$\pm$0.010                          \\
                                 & \multicolumn{2}{c}{GBDT}             & 0.415$\pm$0.026                     & 0.930$\pm$0.018                 & 0.950$\pm$0.009                          \\
                                 & \multicolumn{2}{c}{SVC}              & 0.536$\pm$0.075                     & 0.881$\pm$0.021                 & 0.895$\pm$0.008                          \\ \hline
\end{tabular}
\label{FDR_DD}
\end{table*}

From Table \ref{tab:Result}, the selection of the construction methods have a great impact on model performance. First consider the rectifier data set, the accuracies of three GNN models with SA-based construction method reach approximately 90$\%$ in average, which is much higher than other methods. This is because the SA-based construction method uses a large amount of prior knowledge to construct a high-quality association graph. Besides, the performance of the three FD models using KNN + GAE method for graph construction is better than that using the KNN method; On the motor data set, the performance of GCN using KNN method is slightly better than that of GCN with KNN + GAE method; As for the TEP data set, both construction methods can achieve favorable fault diagnosis performance. The accuracies of the GCN and the GAT models using KNN + GAE method are better than that using KNN.

Generally speaking, the SA-based construction method is better than other methods. However, in most cases, it is difficult to obtain prior knowledge of system, so KNN-based and KNN + GAE-based construction methods can be used as its replacement for cases where the structure information is unknown. Furthermore, the overall performances of GNNs using KNN + GAE-based method are higher than that of GNNs using the KNN-based method. This may be attributed to the fact that, the usage of GAE realizes the prediction of the association relationships that do not exist in the original graph, which actually improves the quality of the association graph.

\begin{figure*}
\begin{center}
\subfigure[Accuracy of GCN model]{
\includegraphics[width=0.3\textwidth]{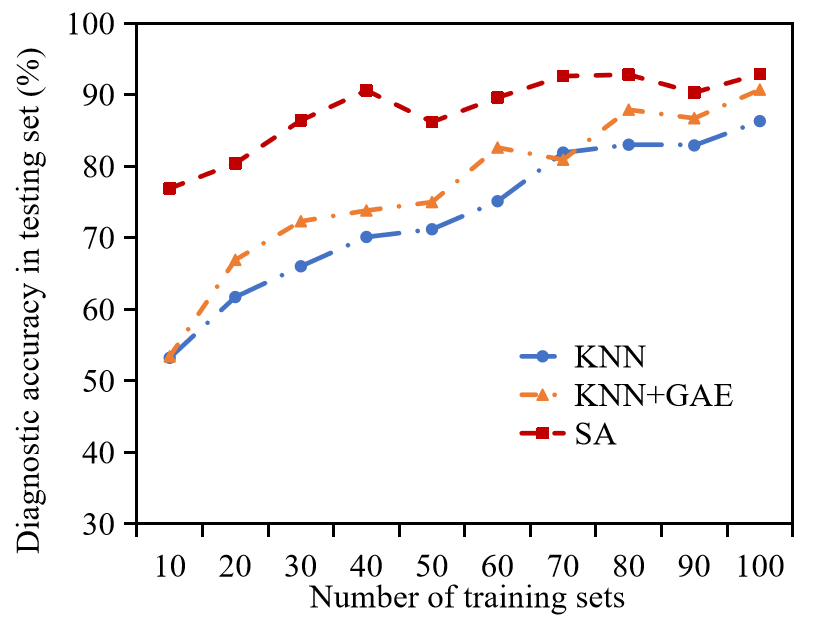}
\label{fig:q_time_all}
}
\subfigure[Accuracy of GAT Model]{
\includegraphics[width=0.3\textwidth]{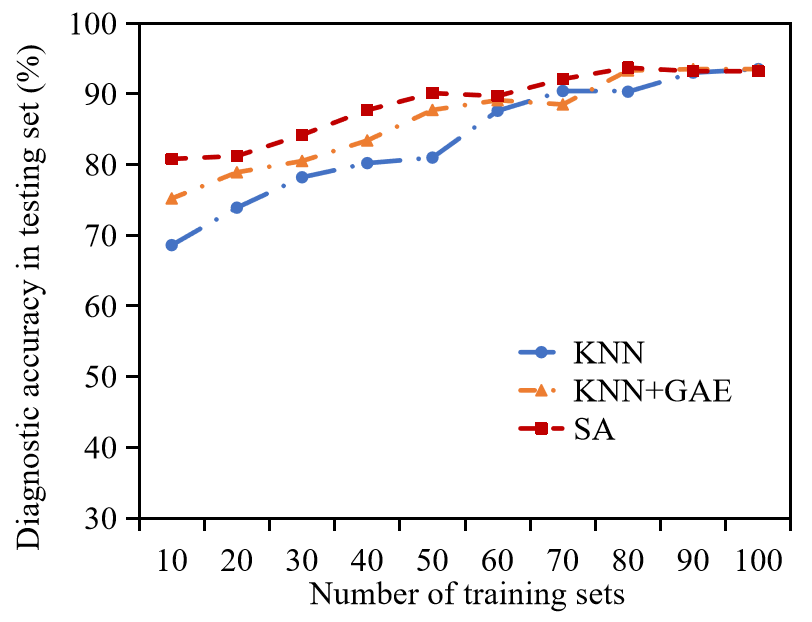}
\label{fig:q_time_sat}
}
\\
\subfigure[Accuracy of GraphSage Model]{
\includegraphics[width=0.3\textwidth]{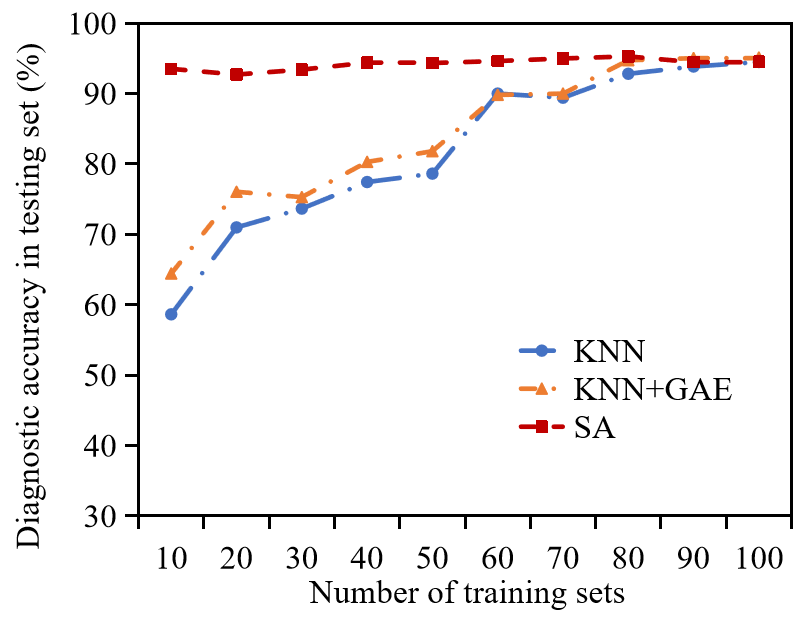}
\label{fig:q_time_unsat}
}
\subfigure[Accuracy of STGCN and ML Models]{
\includegraphics[width=0.3\textwidth]{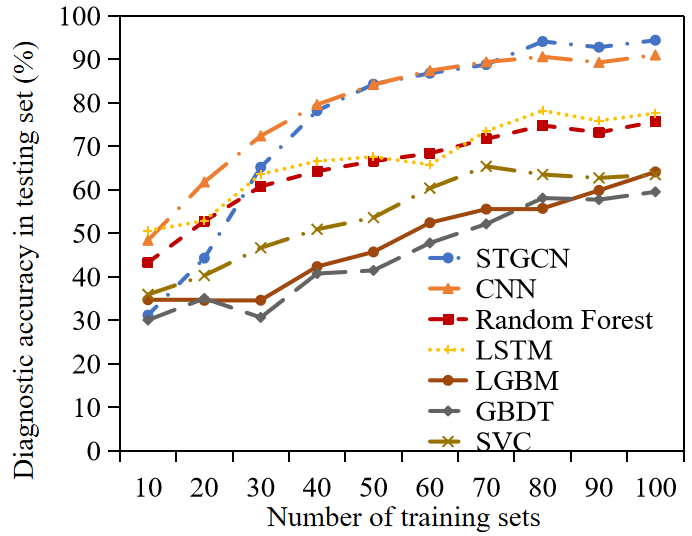}
\label{fig:q_time_sample}
}
\caption{Detailed display of experimental results on the rectifier data set}\label{fig_10}
\end{center}
\end{figure*}

\subsubsection{Comparison between GNN models and baselines}

In the benchmark study, GNN-based methods are also compared with various baseline methods. Now that the SA-based graph construction method introduces prior knowledge, which is infeasible in baseline methods, GNN models trained by SA are not compared with baseline methods. It is found that GCN and GraphSage are not as good as CNN when dealing with the rectifier data set, but achieve better performance than the other baseline methods. STGCN with KNN-based method and GAT with KNN + GAE-based method outperform baseline methods in terms of diagnostic accuracy.

It should be noted that on the motor data set, the LGBM and GBDT methods outperform GNN-based methods as well as other baseline models, this may be because of the fact that the motor data set itself is relatively simple and the association between data is not close.

In general, considering the performance of GNN methods in three data set, GNN methods achieve same or better diagnosis performance compared with the six baseline methods. Besides, when dealing with data set whose association graph can be constructed based on prior knowledge, GNN with SA-based graph construction method is the preferred one.

\begin{table*}[t]
  \centering
  \caption{\label{tab:Comtest}  Comparison results on the rectifier data set}
    \begin{tabular}[H]{ c  c | c  c  c  c  c  c  c  c  c  c}
        \hline

        \multicolumn{2}{c|}{Training set}      &   10   &   20   &   30   &   40   &   50   &   60   &   70   &   80   &   90   &  100  \\
        \hline
        \rule{0pt}{10pt}
                            &   KNN   & 0.586  & 0.710  & 0.736  & 0.774  & 0.786  & 0.900  & 0.894  & 0.928  & 0.938  & 0.945 \\
            GraphSage       & KNN+GAE & 0.644  & 0.761  & 0.753  & 0.803  & 0.818  & 0.898  & 0.897  & 0.947  & \textbf{0.950}  & 0.951 \\
                            &    SA   & \textbf{0.935}  & \textbf{0.927}  & \textbf{0.934}  & \textbf{0.944}  & \textbf{0.943}  & \textbf{0.946}  & \textbf{0.950}  & \textbf{0.953}  & 0.945  & 0.945 \\
        \rule{0pt}{10pt}
                            &   KNN   & 0.532  & 0.617  & 0.660  & 0.701  & 0.712  & 0.751  & 0.819  & 0.830  & 0.829  & 0.863 \\
                GCN         & KNN+GAE & 0.534  & 0.669  & 0.723  & 0.738  & 0.750  & 0.826  & 0.809  & 0.879  & 0.867  & 0.907 \\
                            &   SA    & 0.769  & 0.804  & 0.864  & 0.906  & 0.862  & 0.896  & 0.926  & 0.928  & 0.903  & 0.929 \\
        \rule{0pt}{10pt}
                            &   KNN   & 0.686  & 0.739  & 0.782  & 0.802  & 0.801  & 0.876  & 0.904  & 0.903  & 0.930  & 0.935 \\
                GAT         & KNN+GAE & 0.752  & 0.789  & 0.805  & 0.834  & 0.877  & 0.891  & 0.885  & 0.933  & 0.935  & 0.934 \\
                            &   SA    & 0.808  & 0.812  & 0.842  & 0.877  & 0.901  & 0.897  & 0.921  & 0.937  & 0.932  & 0.932 \\
        \multicolumn{2}{c|}{STGCN}              & 0.312  & 0.443  & 0.652  & 0.782  & 0.843  & 0.868  & 0.888  & 0.941  & 0.928  & \textbf{0.944} \\

        \multicolumn{2}{c|}{CNN}                & 0.484  & 0.618  & 0.724  & 0.796  & 0.842  & 0.874  & 0.894  & 0.906  & 0.893  & 0.910 \\

        \multicolumn{2}{c|}{LSTM}               & 0.506  & 0.529  & 0.636  & 0.666  & 0.676  & 0.658  & 0.735  & 0.782  & 0.759  & 0.776 \\

        \multicolumn{2}{c|}{Random Forest}      & 0.433  & 0.528  & 0.608  & 0.643  & 0.666  & 0.684  & 0.717  & 0.748  & 0.732  & 0.757 \\

        \multicolumn{2}{c|}{LGBT}               & 0.347  & 0.346  & 0.346  & 0.424  & 0.457  & 0.524  & 0.556  & 0.557  & 0.599  & 0.641 \\

        \multicolumn{2}{c|}{GBDT}               & 0.301  & 0.350  & 0.307  & 0.407  & 0.415  & 0.478  & 0.522  & 0.581  & 0.577  & 0.595 \\

        \multicolumn{2}{c|}{SVC}                & 0.360  & 0.403  & 0.467  & 0.509  & 0.536  & 0.604  & 0.654  & 0.635  & 0.627  & 0.635 \\

        \hline
        \rule{0pt}{15pt}
    \end{tabular}
\end{table*}

\begin{figure*}[t]
\centerline{\includegraphics[scale=0.51]{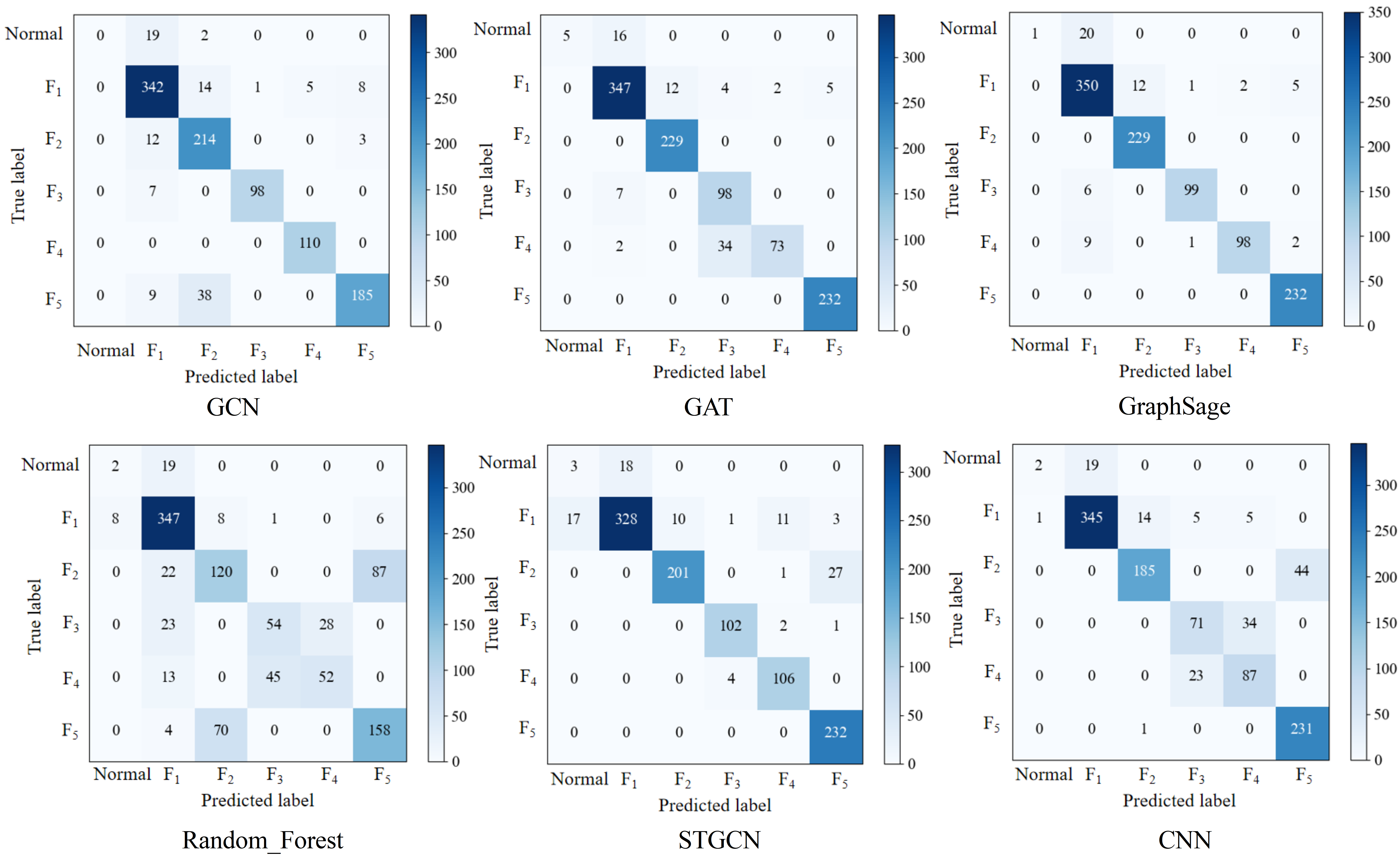}}
\caption{Confusion matrices of various FD methods}\label{fig_13}
\end{figure*}

\begin{figure*}
\begin{center}

\centerline{\includegraphics[scale=0.365]{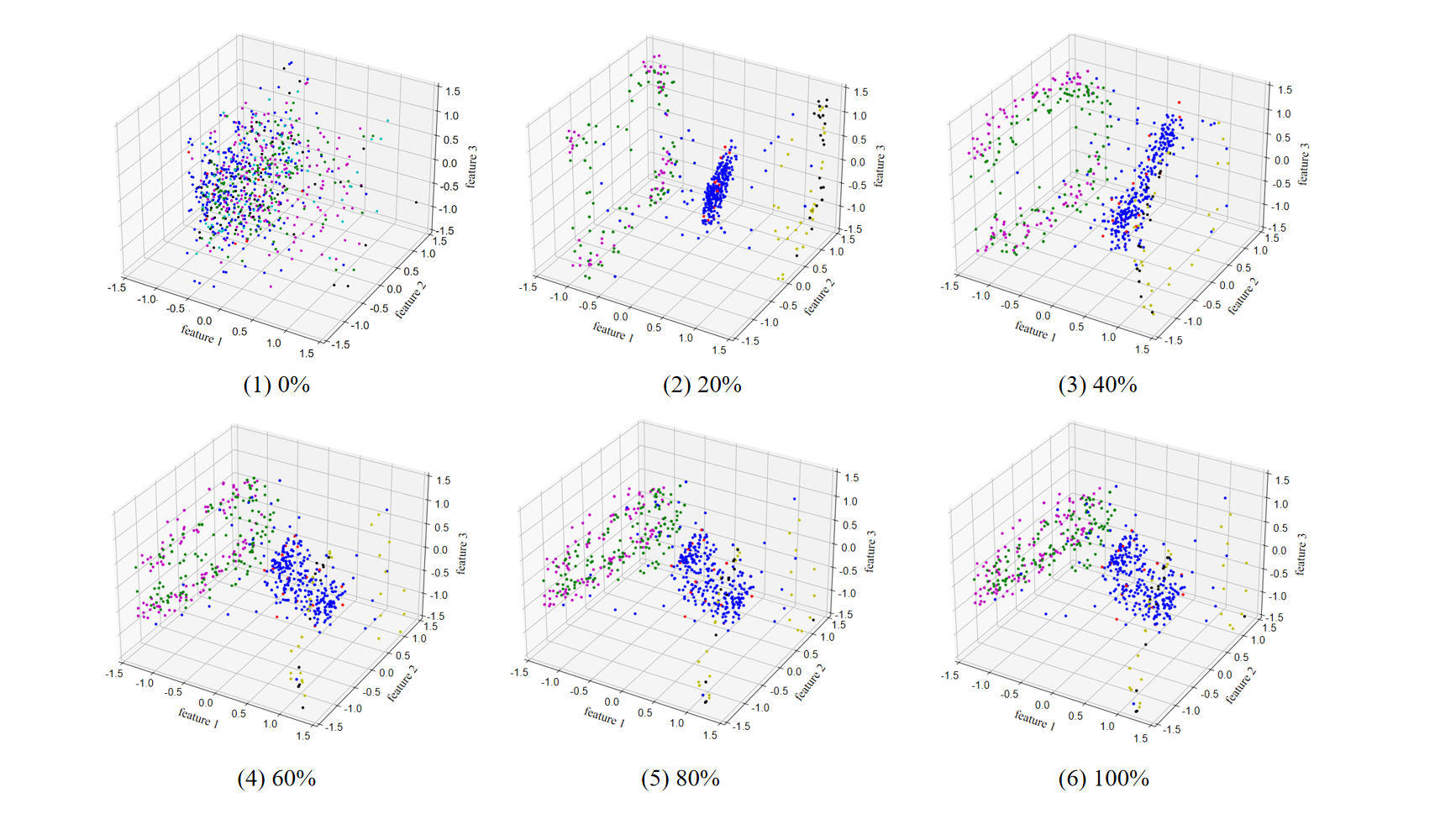}}
\caption{Variation tendency of the hidden layer output of the STGCN model (the percentage represents the proportion of the total iterative training process, and different colors of data denote different categories)}\label{fig_12}

\end{center}
\end{figure*}

\subsubsection{The influence of small sample case}

In practice, a system usually runs in ``long-term fault-free and short term fault'' scenario due to the highly reliable design of the system. This leads to a small number of fault samples, which will be called the ``small sample case'' in this paper.

\begin{figure}[!tb]
\centering
\includegraphics[scale=0.352]{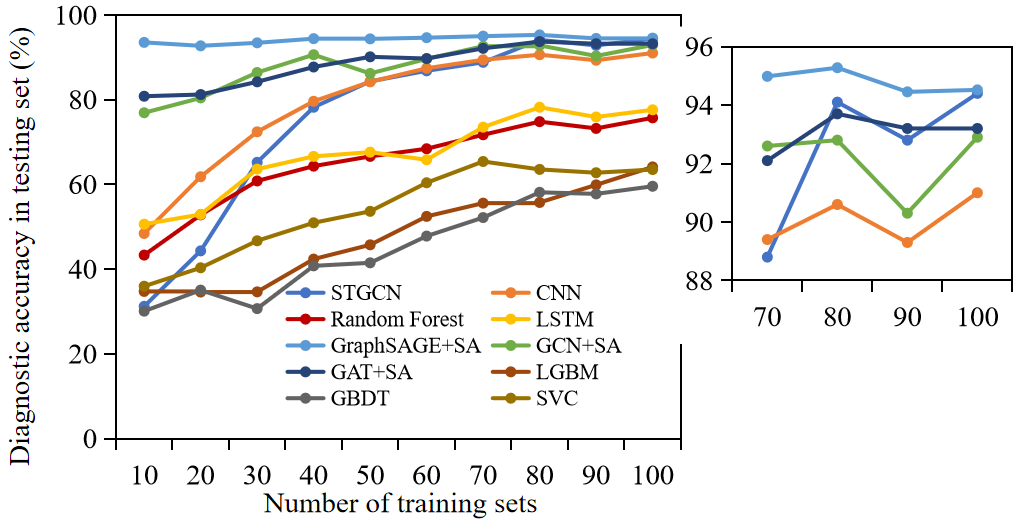}
\caption{Summary of experimental results on the rectifier data set}\label{fig_11}
\end{figure}

To verify the effectiveness of the GNN-based FD methods for the small sample case, the amount of training set is set to 10-100 on the rectifier data set, and a total of ten groups of experiments were conducted for the comparison with baseline methods. The experimental results are shown in Table \ref{tab:Comtest}, Fig. \ref{fig_10}, and Fig. \ref{fig_11}. In order to better present the convergence process inside the STGCN model during its iteration process, the variation tendency of the output of the hidden layer in STGCN method is tracked as shown in Fig. \ref{fig_12}. In this figure, the output of the hidden layer is processed with PCA method for dimensional reduction, and constitutes a node in the graph. As shown in Fig. \ref{fig_12}, with the increase in the number of iterations, the outputs with the same label in the hidden layer become gradually similar. In addition, the confusion matrices corresponding to the diagnosis results obtained by these methods are shown in Fig. \ref{fig_13}. According to the analysis of Table \ref{tab:Comtest}, when the number of training sets is 10 (accounts for 0.94$\%$ of the data set), except for STGCN, the overall diagnostic accuracies of the GNN-based methods are better than the six baseline methods, this may be attributed to the fact that the adjacency matrix of STGCN cannot fully reflect the effective association between features in the case of small samples.

As the size of the training sets increases, the diagnostic accuracies of the GNN-based and baseline FD methods in the test set increase correspondingly, but the diagnostic performance of the GNNs is still better than the methods on the whole. In general, GNN-based methods can achieve relatively better results in the small sample case, and the advantages of the GNNs is more obvious with the introduction of prior knowledge. 

In conclusion, the four GNN-based FD methods have distinct advantages in three benchmark data sets. Most of the methods can achieve good fault diagnosis. Considering that GraphSage belongs to the inductive learning mode, it is a good choice to use GraphSage model for fault diagnosis. GAT uses attention mechanism and a multihead mechanism to improve the generalizability, and has obvious advantages in processing data set with complex internal structures. Compared with the other three GNN-based FD methods, the performance of STGCN is relatively poor when the number of training sets is small, but with the increase of the number of training sets, the accuracy of STGCN in the test sets is approximately equal to GraphSage.

\section{Perspectives of future research}

GNN is a promising way for fault diagnosis, but several challenges still need to be further investigated, which are expected to inspire possible research directions in this field over the next years.

\hangafter 1
\hangindent 1.7em
\noindent
(1) How to construct high quality association graph from data?\\
Up to now, the existing graph construction methods are insufficient, and the data may have poor quality with outliers, uncertain connections, and missing values. Considering that the quality of the graph has a large impact on the performance of GNN model, it is quiet essential to consider how to construct high-quality association graphs from a given data set.

\hangafter 1
\hangindent 1.7em
\noindent
(2) How to use prior knowledge in GNN for fault diagnosis?\\
Most of the data-driven fault diagnosis methods do not pay attention to the prior knowledge of the system of interest, but in many cases, engineers have at least some understanding of the process information. How to combine prior knowledge with measurements is worth considering. A promising way is to combine knowledge graphs with GNN \cite{KG_GNN}.

\hangafter 1
\hangindent 1.7em
\noindent
(3) How to study fault diagnosability?\\
In the field of traditional model-based fault diagnosis, fault diagnosability is divided into two parts: detectability and isolability. The fault that does not meet the requirements of detectability or isolability cannot be diagnosed by fault diagnosis methods. However, in GNN-based fault diagnosis, it is usually assumed that all faults are fully diagnosable, but this assumption does not always hold. In order to get a better diagnosis accuracy, researchers often devote effort to improve the neural network model, but ignore the analysis of the essential problem of fault diagnosability. From this point of view, it is of great significance to analyze fault diagnosability in GNN-based fault diagnosis.

\hangafter 1
\hangindent 1.7em
\noindent
(4) How to update the graph?\\
It is often assumed that the graph used by GNN is unchanged. Can the graph be updated in the process of fault diagnosis so that it can dynamically show the relationship of each node? In addition, the normal status of a system may drift over a period of time. As a result, the association relationship between measurements may also change, which requires timely changes in the association graph.

\hangafter 1
\hangindent 1.7em
\noindent
(5) How to detect unknown fault types?\\
The current fault diagnosis methods are based on known types of faults. When a GNN model receives measurements of an unknown fault type, it makes a fault decision based on the known fault types. The popular transfer learning approach may be a promising technique to provide a feasible solution \cite{Lei_2020_review}.

\hangafter 1
\hangindent 1.7em
\noindent
(6) How to improve GNN security against cyber attacks?\\
In the information age, everything is interconnected. A mature neural network must not only collect data through the internet, but also guard against possible cyber attacks. In addition, GNN model emphasizes the relationship between measurements, so once a small part of measurements in the data set is maliciously tampered with, it will have a negative impact on the performance. How to minimize the above adverse effects is a problem that GNN must face in the industrial application.

\section{Conclusion}


In this paper, the emerging GNN-based fault diagnosis methods are briefly reviewed. The NN-based fault diagnosis methods are divided based on the data representations in real world, namely, the time-series-based NNFD method, the image-based NNFD method, and the graph-based NNFD method. Then, basic principles and principal architectures of GNN are introduced, with attention to GCN, GAT, GraphSage, GAE, and STGCN according to the different graph analytic tasks. Furthermore, the GNN-based fault diagnosis framework is detailed with focus on building the association graph and designing the GNN models. Experiments on three benchmark data sets were carried out to verify the effectiveness and feasibility of GNN-based FD methods by comparing with several baseline FD methods. Finally, perspectives on the challenges of GNN-based fault diagnosis are discussed. This review is prepared in response to the urgent need of a seamless and rigorous transition from classical neural network-based fault diagnosis to the alternative fault diagnosis approaches which operate directly on graph-structured data. It is also expected to provide guidelines for future research in this field.

\section*{Acknowledgment}

This work was supported in part by the National Natural Science Foundation of China ($\#$62173349,$\#$U20A20186), in part by the Swiss National Science Foundation project ($\#$200021$\_$172671): “ALPSFORT: A Learning graph-based framework for cyber-physical systems”.

\bibliographystyle{Cite/IEEEtranTIE}
\bibliography{Cite/centralBib_t}

\begin{thebibliography}{10}
\providecommand{\url}[1]{#1}
\csname url@samestyle\endcsname
\providecommand{\newblock}{\relax}
\providecommand{\bibinfo}[2]{#2}
\providecommand{\BIBentrySTDinterwordspacing}{\spaceskip=0pt\relax}
\providecommand{\BIBentryALTinterwordstretchfactor}{4}
\providecommand{\BIBentryALTinterwordspacing}{\spaceskip=\fontdimen2\font plus
\BIBentryALTinterwordstretchfactor\fontdimen3\font minus
  \fontdimen4\font\relax}
\providecommand{\BIBforeignlanguage}[2]{{%
\expandafter\ifx\csname l@#1\endcsname\relax
\typeout{** WARNING: IEEEtran.bst: No hyphenation pattern has been}%
\typeout{** loaded for the language `#1'. Using the pattern for}%
\typeout{** the default language instead.}%
\else
\language=\csname l@#1\endcsname
\fi
#2}}
\providecommand{\BIBdecl}{\relax}
\BIBdecl

\bibitem{ConMo}
S.~Rizzo, G.~Susinni, and F.~Iannuzzo, ``Intrusiveness of power device
  condition monitoring methods: Introducing figures of merit for condition
  monitoring,'' \emph{IEEE Industrial Electronics Magazine}, 2021,
  doi:{10.1109/MIE.2021.3066959}.

\bibitem{Lei_14}
Y.~G. Lei, J.~Lin, M.~J. Zuo, and Z.~J. He, ``Condition monitoring and fault
  diagnosis of planetary gearboxes: A review,'' \emph{Measurement}, vol.~48,
  pp. 292--305, 2014.

\bibitem{Yin_14}
S.~Yin, S.~X. Ding, X.~Xie, and H.~Luo, ``A review on basic data-driven
  approaches for industrial process monitoring,'' \emph{IEEE Transactions on
  Industrial Electronics}, vol.~61, no.~11, pp. 6418--6428, 2014.

\bibitem{GE201716}
Z.~Ge, ``Review on data-driven modeling and monitoring for plant-wide
  industrial processes,'' \emph{Chemometrics and Intelligent Laboratory
  Systems}, vol. 171, pp. 16--25, 2017.

\bibitem{ZHAO_Zhou_2021}
Y.~Zhao, X.~He, J.~Zhang, H.~Ji, D.~Zhou, and M.~G. Pecht, ``Detection of
  intermittent faults based on an optimally weighted moving average t2 control
  chart with stationary observations,'' \emph{Automatica}, vol. 123, p. 109298,
  2021.

\bibitem{LEE2014314}
J.~Lee, F.~Wu, W.~Zhao, M.~Ghaffari, L.~Liao, and D.~Siegel, ``Prognostics and
  health management design for rotary machinery systems-reviews, methodology
  and applications,'' \emph{Mechanical Systems and Signal Processing}, vol.~42,
  no.~1, pp. 314--334, 2014.

\bibitem{Wang_chenTW16}
J.~Wang, F.~Yang, T.~Chen, and S.~L. Shah, ``An overview of industrial alarm
  systems: Main causes for alarm overloading, research status, and open
  problems,'' \emph{IEEE Transactions on Automation Science and Engineering},
  vol.~13, no.~2, pp. 1045--1061, 2016.

\bibitem{AbnMan}
Y.~Dong, Z.~Wu, X.~Du, J.~Zha, and Q.~Yuan, ``Resource abnormal management
  method of unsteady processes of cloud manufacturing services,'' \emph{China
  Mechanical Engineering}, vol.~29, pp. 1193--1200, 2018.

\bibitem{review}
H.~Henao, G.~Capolino, and M.~Cabanas, ``Trends in fault diagnosis for
  electrical machines: A review of diagnostic techniques,'' \emph{IEEE
  Industrial Electronics Magazine}, vol.~8, no.~2, pp. 31--42, 2014.

\bibitem{3}
J.~Qin, Y.~Dong, Q.~Zhu, J.~Wang, and Q.~Liu, ``Bridging systems theory and
  data science: A unifying review of dynamic latent variable analytics and
  process monitoring,'' \emph{Annual Reviews in Control}, vol.~50, pp. 29--48,
  2020.

\bibitem{1}
S.~Ding, \emph{Data-driven Design of Fault Diagnosis and Fault-tolerant Control
  Systems}.\hskip 1em plus 0.5em minus 0.4em\relax London, UK: Springer, 2014.

\bibitem{2}
S.~Ding, \emph{Model-based Fault Diagnosis Techniques: Design Schemes,
  Algorithms, and Tools}.\hskip 1em plus 0.5em minus 0.4em\relax Berlin,
  Germany: Springer, 2008.

\bibitem{Alippi}
C.~Alippi, N.~Stavros, and R.~Manuel, ``Model-free fault detection and
  isolation in large-scale cyber-physical systems,'' \emph{IEEE Transactions on
  Emerging Topics in Computational Intelligence}, vol.~1, no.~1, pp. 61--71,
  2016.

\bibitem{Luo}
H.~Luo, H.~Zhao, and S.~Yin, ``Data-driven design of fog-computing-aided
  process monitoring system for large-scale industrial processes,'' \emph{IEEE
  Transactions on Industrial Informatics}, vol.~14, pp. 4631--4641, 2018.

\bibitem{ACTA1}
P.~M. Papadopoulos, V.~Reppa, M.~M. Polycarpou, and C.~G. Panayiotou,
  ``Scalable distributed sensor fault diagnosis for smart buildings,''
  \emph{IEEE/CAA Journal of Automatica Sinica}, vol.~7, no.~3, pp. 638--655,
  2020.

\bibitem{4}
Y.~Lecun, L.~Bottou, Y.~Bengio, and P.~Haffner, ``Gradient-based learning
  applied to document recognition,'' \emph{Proceedings of the\ IEEE}, vol.~86,
  no.~11, pp. 2278--2324, 1998.

\bibitem{5}
K.~He, X.~Zhang, S.~Ren, and J.~Sun, ``Deep residual learning for image
  recognition,'' in \emph{IEEE Conference on Computer Vision and Pattern
  Recognition}, pp. 770--778, 2016.

\bibitem{6}
J.~C. David and J.~MacKay, ``A practical bayesian framework fro backpropagation
  networks,'' \emph{Neural Computation}, vol.~4, pp. 448--472, 1992.

\bibitem{IFD}
C.~Cheng, J.~Wang, H.~Chen, Z.~Chen, H.~Luo, and P.~Xie, ``A review of
  intelligent fault diagnosis for high-speed trains: Qualitative approaches,''
  \emph{Entropy}, 2020, doi:{10.3390/e23010001}.

\bibitem{KPI}
K.~Zhang, H.~Hao, Z.~Chen, S.~Ding, and K.~Peng, ``A comparison and evaluation
  of key performance indicator-based multivariate statistics process monitoring
  approaches,'' \emph{Journal of Process Control}, vol.~33, pp. 112--126, 2015.

\bibitem{ACTA2}
K.~Zhong, M.~Han, and B.~Han, ``Data-driven based fault prognosis for
  industrial systems: A concise overview,'' \emph{IEEE/CAA Journal of
  Automatica Sinica}, vol.~7, no.~2, pp. 330--345, 2020.

\bibitem{ACTA3}
D.~Zheng, L.~Zhou, and Z.~Song, ``Kernel generalization of multi-rate
  probabilistic principal component analysis for fault detection in nonlinear
  process,'' \emph{IEEE/CAA Journal of Automatica Sinica}, vol.~8, no.~8, pp.
  1465--1476, 2021.

\bibitem{Yuri}
K.~Zhang, Y.~Shardt, Z.~Chen, X.~Yang, S.~Ding, and K.~Peng, ``A kpi-based
  process monitoring and fault detection framework for large-scale processes,''
  \emph{ISA Transactions}, vol.~68, pp. 276--286, 2017.

\bibitem{ChenHT_20_review}
H.~Chen, B.~Jiang, S.~X. Ding, and B.~Huang, ``Data-driven fault diagnosis for
  traction systems in high-speed trains: A survey, challenges, and
  perspectives,'' \emph{IEEE Transactions on Intelligent Transportation
  Systems}, pp. 1--17, 2020, doi:{10.1109/TITS.2020.3029946}.

\bibitem{10-}
S.~Kiranyaz, O.~Avci, O.~Abdeljaber, T.~Ince, M.~Gabbouj, and D.~Inman, ``1d
  convolutional neural networks and applications: A survey,'' \emph{Mechanical
  Systems and Signal Processing}, vol. 151, p. 107398, 2021,
  doi:{10.1016/j.ymssp.2020.107398}.

\bibitem{11-}
M.~Kuppusamy, A.~Hussain, P.~Sanjeevikumar, J.~Holm-Nielsen, and V.~Kaliappan,
  ``Deep learning for fault diagnostics in bearings, insulators, pv panels,
  power lines, and electric vehicle applications-the state-of-the-art
  approaches,'' \emph{IEEE Access}, vol.~9, pp. 41\,246--41\,260, 2021.

\bibitem{FD_Survey}
D.~Hoang and H.~Kang, ``A survey on deep learning based bearing fault
  diagnosis,'' \emph{Neurocomputing}, vol. 335, pp. 327--335, 2019.

\bibitem{43}
T.~Kipf and M.~Welling, ``Semi-supervised classification with graph
  convolutional networks,'' \emph{International Conference on Learning
  Representations}, vol. abs/1609.02907, 2016.

\bibitem{14}
H.~Zhang and Y.~Shen, ``Template-based prediction of protein structure with
  deep learning,'' \emph{BMC Genomics}, vol.~21, 2020.

\bibitem{15}
M.~Defferrard, X.~Bresson, and P.~Vandergheynst, ``Convolutional neural
  networks on graphs with fast localized spectral filtering,'' \emph{eural
  information processing systems}, vol.~30, pp. 3844--3852, 2016.

\bibitem{16}
M.~Niepert, M.~Ahmed, and K.~Kutzkov, ``Learning convolutional neural networks
  for graphs,'' \emph{International Conference on Machine Learning}, pp.
  2014--2023, 2016.

\bibitem{R-CNN}
R.~Girshick, J.~Donahue, and T.~Darrell, ``Rich feature hierarchies for
  accurate object detection and semantic segmentation,'' \emph{Computer Vision
  and Pattern Recognition}, pp. 580--587, 2014.

\bibitem{17}
X.~Qi, R.~Liao, J.~Jia, S.~Fidler, and R.~Urtasun, ``3{D} graph neural networks
  for {RGBD} semantic segmentation,'' pp. 5209--5218, 2017.

\bibitem{18}
L.~Yao, C.~Mao, and Y.~Luo, ``Graph convolutional networks for text
  classification,'' \emph{Proceedings of the AAAI Conference on Artificial
  Intelligence}, vol.~33, pp. 7370--7377, 2019.

\bibitem{19}
N.~Park, A.~Kan, X.~L. Dong, T.~Zhao, and C.~Faloutsos, ``Estimating node
  importance in knowledge graphs using graph neural networks,'' in
  \emph{International Conference on Knowledge Discovery and Data Mining}, pp.
  596--606, 2019.

\bibitem{GNNSur1}
Z.~Zhang, P.~Cui, and W.~Zhu, ``Deep learning on graphs: A survey,'' \emph{IEEE
  Transactions on Knowledge and Data Engineering}, 2020, doi:
  {10.1109/TKDE.2020.2981333}.

\bibitem{GNNSur2}
I.~Chami, S.~Abu, and B.~Perozzi, ``Machine learning on graphs: A model and
  comprehensive taxonomy,'' \emph{CoRR}, 2020.

\bibitem{GNNSur3}
D.~Bacciu, F.~Errica, A.~Micheli, and M.~Podda, ``A gentle introduction to deep
  learning for graphs,'' \emph{Neural Networks}, vol. 129, pp. 203--221, 2020.

\bibitem{20}
J.~Zhou, G.~Cui, and Z.~Zhang, ``Graph neural networks: A review of methods and
  applications,'' 2018.

\bibitem{22}
J.~B. Lee, R.~A. Rossi, S.~Kim, N.~K. Ahmed, and E.~Koh, ``Attention models in
  graphs: A survey,'' \emph{ACM Transactions on Knowledge Discovery from Data},
  vol.~13, pp. 1--25, 2019.

\bibitem{21}
Z.~Wu, S.~Pan, F.~Chen, G.~Long, C.~Zhang, and P.~Yu, ``A comprehensive survey
  on graph neural networks,'' \emph{IEEE Transactions on Neural Networks and
  Learning Systems}, vol.~32, no.~1, pp. 4--24, 2021.

\bibitem{23}
K.~Chen, J.~Hu, Y.~Zhang, Z.~Yu, and J.~He, ``Fault location in power
  distribution systems via deep graph convolutional networks,'' \emph{IEEE
  Journal on Selected Areas in Communications}, vol.~38, pp. 119--131, 2019.

\bibitem{24}
J.~Jiang, J.~Chen, and T.~Gu, ``Anomaly detection with graph convolutional
  networks for insider threat and fraud detection,'' in \emph{IEEE Military
  Communications Conference}, pp. 109--114, 2019.

\bibitem{63}
X.~Yu, B.~Tang, and K.~Zhang, ``Fault diagnosis of wind turbine gearbox using a
  novel method of fast deep graph convolutional networks,'' \emph{IEEE
  Transactions on Instrumentation and Measurement}, vol.~70, pp. 1--14, 2021.

\bibitem{25}
T.~Li, Z.~Zhao, C.~Sun, R.~Yan, and X.~Chen, ``Multi-receptive field graph
  convolutional networks for machine fault diagnosis,'' \emph{IEEE Transactions
  on Industrial Electronics}, 2020, doi: {10.1109/TIE.2020.3040669}.

\bibitem{26}
C.~Li, L.~Mo, and R.~Yan, ``Rolling bearing fault diagnosis based on horizontal
  visibility graph and graph neural networks,'' in \emph{International
  conference on Sensing, measurement, data analytics in the era of artificial
  intelligence}, pp. 275--279, 2020.

\bibitem{35}
W.~Z. A.~J.~Bruna and Y.~L, ``Spectral networks and locally connected networks
  on graphs,'' \emph{Computing Research Repository}, pp. 1--14, 2013.

\bibitem{36}
Q.~Li, Z.~Han, and X.~Wu, ``Deeper insights into graph convolutional networks
  for semi-supervised learning,'' in \emph{Proceedings of the Thirty-Second
  {AAAI} Conference on Artificial Intelligence, (AAAI-18), New Orleans,
  Louisiana, USA, February 2-7, 2018}, S.~A. McIlraith and K.~Q. Weinberger,
  Eds., pp. 3538--3545.\hskip 1em plus 0.5em minus 0.4em\relax {AAAI} Press,
  2018.

\bibitem{37}
Z.~Zhang, J.~Huang, and Q.~Tan, ``{SR-HGAT}: Symmetric relations based
  heterogeneous graph attention network,'' \emph{IEEE Access}, vol.~8, pp.
  631--645, 2020.

\bibitem{Deepwalk}
H.~Li, H.~Chen, and W.~Wang, ``A structural deep network embedding model for
  predicting associations between mirna and disease based on molecular
  association network,'' \emph{Scientific Reports}, vol.~11, p. 12640, 2021.

\bibitem{Graphclu}
A.~Tsitsulin, J.~Palowitch, B.~Perozzi, and E.~Muller, ``Graph clustering with
  graph neural networks,'' \emph{ArXiv}, vol. abs/2006.16904, 2020.

\bibitem{32}
Z.~Wang and T.~Oates, ``Imaging time-series to improve classification and
  imputation,'' \emph{Conference on artificial intelligence}, pp. 3939--3945,
  2015.

\bibitem{52}
Z.~Wu, S.~Pan, F.~Chen, G.~Long, C.~Zhang, and S.~Philip, ``A comprehensive
  survey on graph neural networks,'' \emph{IEEE Transactions on Neural Networks
  and Learning Systems}, vol.~32, no.~1, pp. 1--21, 2020.

\bibitem{58}
P.~Velickovic, G.~Cucurull, A.~Casanova, A.~Romero, and Y.~Bengio, ``Graph
  attention networks,'' \emph{International Conference on Learning
  Representations}, 2017.

\bibitem{59}
J.~He and H.~Zhao, ``Fault diagnosis and location based on graph neural network
  in telecom networks,'' in \emph{International Conference on Networking and
  Network Applications}, pp. 304--309, 2020.

\bibitem{60}
W.~Hamilton, R.~Ying, and J.~Leskovec, ``Inductive representation learning on
  large graphs,'' in \emph{Neural Information Processing Systems}, vol.~30, pp.
  1025--1035, 2017.

\bibitem{55}
T.~N. Kipf and M.~Welling, ``Variational graph auto-encoders,'' 2016.

\bibitem{61}
S.~Pan, R.~Hu, G.~Long, J.~Jiang, and L.~Yao, ``Adversarially regularized graph
  autoencoder for graph embedding,'' in \emph{Proceedings of the International
  Joint Conference on Artificial Intelligence}, pp. 2609--2615, 2018.

\bibitem{62}
Y.~Liao, Y.~Wang, and Y.~Liu, ``Graph regularized auto-encoders for image
  representation,'' \emph{IEEE Transactions on Image Processing}, vol.~26,
  no.~6, pp. 2839--2852, 2016.

\bibitem{54}
B.~Yu, H.~Yin, and Z.~Zhu, ``Spatio-temporal graph convolutional networks: A
  deep learning framework for traffic forecasting,'' in \emph{Proceedingsof the
  International Joint Conference on Artificial Intelligence}, pp. 3634--3640,
  2018.

\bibitem{42}
E.~Mansimov, O.~Mahmood, S.~Kang, and K.~Cho, ``Molecular geometry prediction
  using a deep generative graph neural network,'' \emph{Scientific Reports},
  vol.~9, no.~1, pp. 1--13, 2019.

\bibitem{45}
T.~Li, Z.~Zhao, C.~Sun, R.~Yan, and X.~Chen, ``Multi-receptive field graph
  convolutional networks for machine fault diagnosis,'' \emph{IEEE Transactions
  on Industrial Electronics}, 2020, doi: {10.1109/TIE.2020.3040669}.

\bibitem{46}
L.~Franceschi, M.~Niepert, M.~Pontil, and X.~He, ``Learning discrete structures
  for graph neural networks,'' \emph{International Conference on Machine
  Learning}, 2019.

\bibitem{47}
Z.~Chen, J.~Xu, T.~Peng, and C.~Yang, ``Graph convolutional network-based
  method for fault diagnosis using a hybrid of measurement and prior
  knowledge,'' \emph{IEEE Transactions on Cybernetics}, pp. 1--13, 2021, doi:
  10.1109/TCYB.2021.3059002.

\bibitem{48}
R.~Berg, T.~Kipf, and M.~Welling, ``Graph convolutional matrix completion,''
  \emph{ArXiv}, vol. abs/1706.02263, 2017.

\bibitem{49}
Y.~Hou, J.~Zhang, J.~Cheng, and K.~Ma, ``Measuring and improving the use of
  graph information in graph neural networks,'' in \emph{International
  Conference on Learning Representations}, 2019.

\bibitem{YC}
C.~Yang, C.~Yang, T.~Peng, X.~Yang, and W.~Gui, ``A fault-injection strategy
  for traction drive control systems,'' \emph{IEEE Transactions on Industrial
  Electronics}, vol.~64, no.~7, pp. 5719--5727, 2017.

\bibitem{KG_GNN}
S.~Arora, ``A survey on graph neural networks for knowledge graph completion,''
  \emph{CoRR}, vol. abs/2007.12374, 2020.

\bibitem{Lei_2020_review}
Y.~Lei, B.~Yang, X.~Jiang, F.~Jia, N.~Li, and A.~K. Nandi, ``Applications of
  machine learning to machine fault diagnosis: A review and roadmap,''
  \emph{Mechanical Systems and Signal Processing}, vol. 138, p. 106587, 2020.

\end{thebibliography}

\end{document}